# Approximate Strong Equilibrium in Job Scheduling Games

**Michal Feldman** MFELDMAN@HUJI.AC.IL
*School of Business Administration*
*and Center for the Study of Rationality,*
*Hebrew University of Jerusalem, Israel.*

**Tami Tamir** TAMI@IDC.AC.IL
*School of Computer Science,*
*The Interdisciplinary Center, Herzliya, Israel.*

## Abstract

A Nash Equilibrium (NE) is a strategy profile resilient to unilateral deviations, and is predominantly used in the analysis of multiagent systems. A downside of NE is that it is not necessarily stable against deviations by coalitions. Yet, as we show in this paper, in some cases, NE does exhibit stability against coalitional deviations, in that the benefits from a joint deviation are bounded. In this sense, NE approximates *strong equilibrium.*

Coalition formation is a key issue in multiagent systems. We provide a framework for quantifying the stability and the performance of various assignment policies and solution concepts in the face of coalitional deviations. Within this framework we evaluate a given configuration according to three measures: (i) $IR_{min}$: the maximal number $\alpha$, such that there exists a coalition in which the minimal improvement ratio among the coalition members is $\alpha$, (ii) $IR_{max}$: the maximal number $\alpha$, such that there exists a coalition in which the maximal improvement ratio among the coalition members is $\alpha$, and (iii) $DR_{max}$: the maximal possible damage ratio of an agent outside the coalition.

We analyze these measures in job scheduling games on identical machines. In particular, we provide upper and lower bounds for the above three measures for both NE and the well-known assignment rule *Longest Processing Time* (LPT).

Our results indicate that LPT performs better than a general NE. However, LPT is not the best possible approximation. In particular, we present a polynomial time approximation scheme (PTAS) for the makespan minimization problem which provides a schedule with $IR_{min}$ of $1 + \varepsilon$ for any given $\epsilon$. With respect to computational complexity, we show that given an NE on $m \geq 3$ identical machines or $m \geq 2$ unrelated machines, it is NP-hard to determine whether a given coalition can deviate such that every member decreases its cost.

## 1. Introduction

We consider job scheduling systems, in which $n$ jobs are assigned to $m$ identical machines and incur a cost which is equal to the total load on the machine they are assigned to.[1] These problems have been widely studied in recent years from a game theoretic perspective (Koutsoupias & Papadimitriou, 1999; Andelman, Feldman, & Mansour, 2007; Christodoulou, Koutsoupias, & Nanavati, 2004; Czumaj & Vöcking, 2002; A. Fiat & Olonetsky., 2007). In contrast to the traditional setting, where a central designer determines the allocation of jobs to machines and all the participating entities are assumed to obey the protocol, mul-

1. This cost function characterizes systems in which jobs are processed in parallel, or when all jobs on a particular machine have the same single pick-up time, or need to share some resource simultaneously.

tiagent systems are populated by heterogeneous, autonomous agents, which often display selfish behavior. Different machines and jobs may be owned by different *strategic* entities, who will typically attempt to optimize their own objective rather than the global objective. Game theoretic analysis provides us with the mathematical tools to study such situations, and indeed has been extensively used recently to analyze multiagent systems. This trend is motivated in part by the emergence of the Internet, which is composed of distributed computer networks managed by multiple administrative authorities and shared by users with competing interests (Papadimitriou, 2001).

Most game theoretic models applied to job scheduling problems, as well as other network games (e.g., Fabrikant, Luthra, Maneva, Papadimitriou, & Shenker, 2003; Albers, Elits, Even-Dar, Mansour, & Roditty, 2006; Roughgarden & Tardos, 2002; Anshelevich, Dasgupta, Kleinberg, Tardos, Wexler, & Roughgarden, 2004), use the solution concept of *Nash equilibrium* (NE), in which the strategy of each agent is a best response to the strategies of all other agents. While NE is a powerful tool for analyzing outcomes in competitive environments, its notion of stability applies only to unilateral deviations. In numerous multiagent settings, selfish agents stand to benefit from cooperating by forming coalitions (Procaccia & Rosenschein, 2006). Therefore, even when no single agent can profit by a unilateral deviation, NE might still not be stable against a group of agents *coordinating* a joint deviation, which is profitable to *all the members* of the group. This stronger notion of stability is exemplified in the *strong equilibrium* (SE) solution concept, coined by Aumann (1959). In a strong equilibrium, no coalition can deviate and improve the utility of *every* member of the coalition.

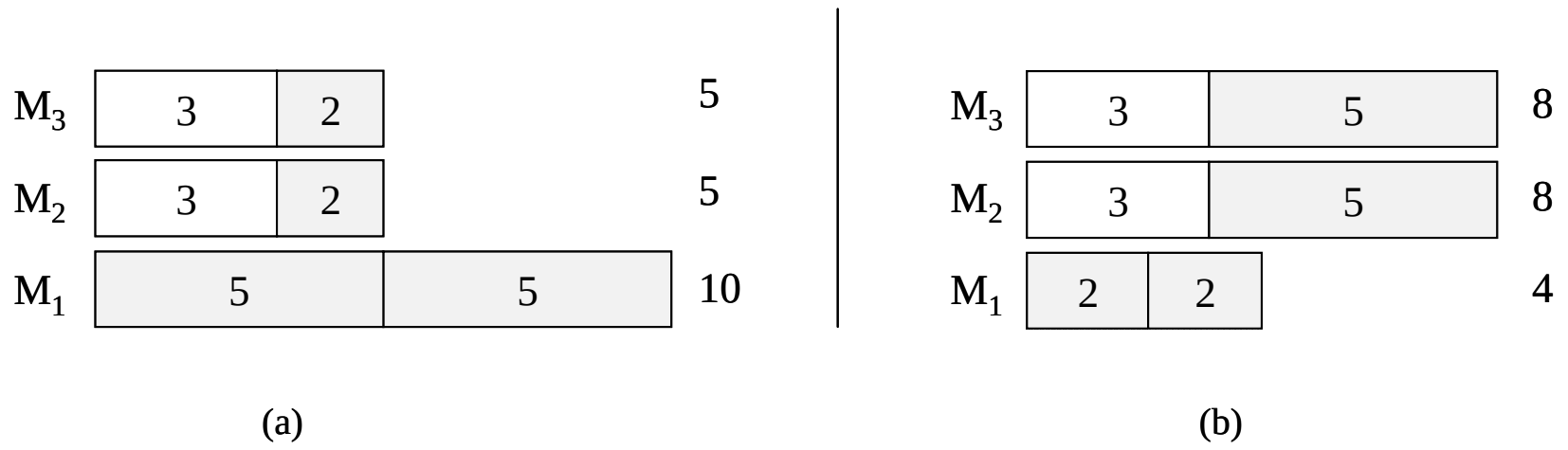

Figure 1: An example of a configuration (a) that is a Nash equilibrium but is not resilient against coordinated deviations, since the jobs of load 5 and 2 all profit from the deviation demonstrated in (b).

As an example, consider the configuration depicted in Figure 1(a). In our figures, a job is represented by a rectangle whose width corresponds to the job's length. The jobs scheduled on a specific machine form a vertical concatenation of rectangles. For example, in Figure 1(a) there are three machines, and $M_1$ processes two jobs of length 5. Note that the internal order of jobs has no effect, since the cost of each job is defined to be the load on the machine it is assigned to. This configuration is an NE since no job can reduce its cost through a unilateral deviation. One might think that any NE on identical machines must also be sustainable against joint deviations. Yet, as was already observed in (Andelman et al., 2007), this may not be true.[2] For example, the configuration above is not resilient against a coordinated deviation of the coalition $\Gamma$ consisting of the four jobs of load 5 and 2 deviating to configuration (b), where the jobs of load 5 decrease their costs from 10 to 8, and the jobs of load 2 improve from 5 to 4. Note that the cost of the two jobs of load 3 (which are not members of the coalition) increases.

In the example above, every member of the coalition improves its cost by a (multiplicative) factor of $\frac{5}{4}$. By how much more can a coalition improve? Is there a bound on the *improvement ratio*? As it will turn out, this example is in fact the most extreme one in a sense that will be clarified below. Thus, while NE is not completely stable against coordinated deviations, in some settings, it does provide us with some notion of approximate stability to coalitional deviations (or *approximate strong equilibrium*).

We also consider a subclass of NE schedules, produced by the *Longest Processing Time* (LPT) rule (Graham, 1969). The LPT rule sorts the jobs in a non-increasing order of their loads and greedily assigns each job to the least loaded machine. It is easy to verify that every configuration produced by the LPT rule is an NE (Fotakis, S. Kontogiannis, & Spiraklis, 2002). Is it also an SE? Note that for the instance depicted in Figure 1, LPT would have produced an SE. However, as we show, this is not always the case.

In this paper we provide a framework for studying the notion of approximate stability to coalitional deviations. In our analysis, we consider three different measures. The first two measure the stability of a configuration, and uses the notion of an *improvement ratio* of a job, which is defined as the ratio between the job's costs before and after the deviation. The third measures the worst possible effect on the non-deviating jobs, as will be explained below.[3]

**1. Minimum Improvement Ratio:** By definition, all members of a coalition must reduce their cost. That is, the improvement ratio of every member of the coalition is larger than 1. Clearly, the coalition members might not share the same improvement ratio. The minimum improvement ratio of a particular deviation is the minimal improvement ratio of a coalition member. The minimum improvement ratio of a schedule $s$, denoted $IR_{min}(s)$, is the maximum over all possible deviations originated from $s$ of the minimal improvement ratio of the deviation. In other words, there is no coalitional deviation originating from $s$ such that every member of the coalition reduces its cost by a factor greater than $IR_{min}(s)$.

A closely related notion has been suggested by Albers (2009), who defined a strategy profile to be an $\alpha$-SE if there is no coalition in which each agent can improve by a factor of more than $\alpha$. In our notation, a schedule $s$ is an $\alpha$-SE if $IR_{min}(s)$ is at most $\alpha$. Albers studied this notion in the context of SE existence in cost-sharing games, and showed that for a sufficiently large $\alpha$, an $\alpha$-SE always exists. The justification behind this concept is that agents may be willing to deviate only if they improve by a sufficiently high factor (due to, for example, some overhead associated with the deviation).

2. This statement holds for $m \geq 3$. For 2 identical machines, every NE is also an SE (Andelman et al., 2007).
3. Throughout this paper, we define approximation by a *multiplicative* factor. Since the improvement and damage ratios for all the three measures presented below are constants greater than one (as will be shown below), the *additive* ratios are unbounded. Formally, for any value $a$ it is possible to construct instances (by scaling the instances we provide for the multiplicative ratio) in which the cost of all jobs is reduced, or the cost of some jobs is increased, by at least an additive factor of $a$.

For three machines, we show that every NE is a $\frac{5}{4}$-SE. That is, there is no coalition that can deviate such that every member improves by a factor larger than $\frac{5}{4}$. For this case, we also provide a matching lower bound (recall Figure 1 above), that holds for any $m \geq 3$. For arbitrary $m$, we show that every NE is a $(2 - \frac{2}{m+1})$-SE. Our proof technique draws a connection between *makespan* approximation and approximate stability, where the makespan of a configuration is defined as the maximum load on any machine in the configuration.

We next consider schedules obtained from the LPT rule. We show that for $m = 3$, every LPT configuration is a $(\frac{1}{2} + \frac{\sqrt{6}}{4})$-SE ($\approx 1.1123$), and we also provide a matching lower bound, which holds for any $m \geq 3$. For arbitrary $m$, we show an upper bound of $\frac{4}{3} - \frac{1}{3m}$.

The above results indicate that LPT is more stable than NE with respect to coalitional deviations. Yet, LPT is not the best possible approximation of SE. Similar to this notion in approximation algorithms, we define an SE-PTAS to be an assignment algorithm which gets as input an additional parameter $\varepsilon$, specifying how close to an SE the schedule should be and produces a $(1 + \varepsilon)$-SE in time polynomial in $n, 1/\varepsilon$. In this paper we devise an SE-PTAS for any fixed number of machines, which also approximates the makespan within a factor of $1 + \epsilon$.

**2. Maximum Improvement Ratio:** The maximum improvement ratio of a particular deviation is the maximal improvement ratio experienced by some coalition member. The maximum improvement ratio of a schedule $s$, denoted $IR_{max}(s)$, is the maximum over all possible deviations originated from $s$ of the maximal improvement ratio of the deviation. In other words, there is no coalition deviation originating from $s$ such that there exists a member of the coalition that reduces its cost by a factor greater than $IR_{max}(s)$.

This notion establishes the bounds on how much an agent would gain in a deviating coalition for which all agents gain something from the deviation. Also, this notion is similar in spirit to stability against a large *total* improvement. It also suits environments in which individuals are willing to obey a specific player as long as they are not hurt. Interestingly, we find that given an NE configuration, the improvement ratio of a single agent may be arbitrarily large, for any $m \geq 3$. In contrast, for LPT configurations on three machines, no agent can improve by a factor of $\frac{5}{3}$ or more and this bound is tight. Thus, with respect to $IR_{max}$, the relative stability of LPT compared to NE is more significant than with respect to $IR_{min}$. For arbitrary $m$, we provide a lower bound of $2 - \frac{1}{m}$, which we believe to be tight.

**3. Maximum Damage Ratio:** As is the case for the jobs of load 3 in Figure 1, some jobs might get hurt as a result of a coalitional deviation. The third measure that we consider is the worst possible effect of a deviation on jobs that are not members of the deviating coalition. Formally, the *maximum damage ratio* is the maximal ratio between the costs of a non-coalition member before and after the deviation. Variants of this measure have been considered in distributed systems, e.g., the Byzantine Generals problem (Lamport, Shostak, & Pease, 1982), and in rational secret sharing (Halpern & Teague, 2004).[4] In Section 5, we prove that the maximum damage ratio is less than 2 for any NE configuration, and less

4. In a rational secret sharing protocol, a set of players, each holding a share of a secret, aims to jointly reconstruct it. Viewing the protocol as a game, the players' utilities are typically assumed to satisfy the following two basic constraints: ($i$) each player prefers learning the secret over not learning it, and ($ii$) conditioned on having learned the secret, each player prefers as few as possible other players to learn it.

than $\frac{3}{2}$ for any LPT configuration. Both bounds hold for any $m \geq 3$, and for both cases we provide matching lower bounds.

In summary, our results in Sections 3-5 (see Table 1) indicate that NE configurations are approximately stable with respect to the $IR_{min}$ measure. Moreover, the performance of jobs outside the coalition would not be hurt by much as a result of a coalitional deviation. As for $IR_{max}$, our results provide an additional strength of the LPT rule, which is already known to possess attractive properties (with respect to, e.g., makespan approximation and stability against unilateral deviations).

| | $IR_{min}$ | | | $IR_{max}$ | | $DR_{max}$ | |
|---|---|---|---|---|---|---|---|
| | upper bound | | lower bound | upper bound | lower bound | upper bound | lower bound |
| | $m \geq 3$ | $m = 3$ | | | | | |
| NE | $2 - \frac{2}{m+1}$ | $\frac{5}{4}$ | $\frac{5}{4}$ | unbounded | | 2 | 2 |
| LPT | $\frac{4}{3} - \frac{1}{3m}$ | $\frac{1}{2} + \frac{\sqrt{6}}{4}$ | $\frac{1}{2} + \frac{\sqrt{6}}{4}$ | $\frac{5}{3}$ $(m{=}3)$ | $2 - \frac{1}{m}$ | $\frac{3}{2}$ | $\frac{3}{2}$ |

Table 1: Our results for the three measures. Unless specified otherwise, the results hold for arbitrary number of machines $m$.

In Section 7, we study computational complexity aspects of coalitional deviations. We find that it is NP-hard to determine whether an NE configuration on $m \geq 3$ identical machines is an SE. Moreover, given a particular configuration and a set of jobs, it is NP-hard to determine whether this set of jobs can engage in a coalitional deviation. For unrelated machines (i.e., where each job incurs a different load on each machine), the above hardness results hold already for $m = 2$ machines. These results might have implications on coalitional deviations with computationally restricted agents.

**Related work:** NE is shown in this paper to provide approximate stability against coalitional deviations. A related body of work studies how well NE approximates the optimal outcome of competitive games. The Price of Anarchy was defined as the ratio between the worst-case NE and the optimum solution (Papadimitriou, 2001; Koutsoupias & Papadimitriou, 1999), and has been extensively studied in various settings, including job scheduling (Koutsoupias & Papadimitriou, 1999; Christodoulou et al., 2004; Czumaj & Vöcking, 2002), network design (Albers et al., 2006; Anshelevich et al., 2004; Anshelevich, Dasgupta, Tardos, Wexler, & Roughgarden, 2003; Fabrikant et al., 2003), network routing (Roughgarden & Tardos, 2002; Awerbuch, Azar, Richter, & Tsur, 2003; Christodoulou & Koutsoupias, 2005), and more.

The notion of strong equilibrium (SE) (Aumann, 1959) expresses stability against coordinated deviations. The downside of SE is that most games do not admit any SE, even amongst those admitting a Nash equilibrium. Various recent works have studied the existence of SE in particular families of games. For example, it has been shown that in every job scheduling game and (almost) every network creation game, an SE exists (Andelman et al., 2007). In addition, several papers (Epstein, Feldman, & Mansour, 2007; Holzman & Law-Yone, 1997, 2003; Rozenfeld & Tennenholtz, 2006) provided a topological characterization for the existence of SE in different congestion games, including routing and cost-sharing connection games. The vast literature on SE (e.g., Holzman & Law-Yone, 1997, 2003; Milchtaich, 1998; Bernheim, Peleg, & Whinston, 1987) concentrate on pure strategies and pure deviations, as is the case in our paper. In job scheduling settings, it was shown by Andelman et al. (2007) that if mixed deviations are allowed, it is often the case that no SE exists. When an SE exists, clearly, the price of anarchy with respect to SE (denoted the *strong price of anarchy* by Andelman et al., 2007) is significantly better than the price of anarchy with respect to NE (Andelman et al., 2007; A. Fiat & Olonetsky., 2007; Leonardi & Sankowski, 2007).

Following our work, $IR_{min}$ bounds for the case of $m = 4$ machines have been provided by Chen (2009), who extended our bound of $\frac{5}{4}$ for NE schedules, and provided a bound of $(\frac{1}{2} + \frac{\sqrt{345}}{30}) \approx 1.119$ for LPT-based schedules.

## 2. Model and Preliminaries

In this section we give a formal description of the model and provide several useful observations and properties of deviations by coalitions.

### 2.1 Resilience to Deviations by Coalitions

We first present a general game theoretic setting and then describe the specific job scheduling setting which is the focus of this paper.

A game is denoted by a tuple $G = \langle N, (S_j), (c_j) \rangle$, where $N = \{1, \ldots, n\}$ is the set of players, $S_j$ is the finite action space of player $j \in N$, and $c_j$ is the cost function of player $j$. The joint action space of the players is $S = \times_{i=1}^{n} S_i$. For a joint action $s = (s_1, \ldots, s_n) \in S$, we denote by $s_{-j}$ the actions of players $j' \neq j$, i.e., $s_{-j} = (s_1, \ldots, s_{j-1}, s_{j+1}, \ldots, s_n)$. Similarly, for a set of players $\Gamma$, also called a *coalition*, we denote by $s_\Gamma$ and $s_{-\Gamma}$ the actions of players in $\Gamma$ and not in $\Gamma$, respectively. The cost function of player $j$ maps a joint action $s \in S$ to a real number, i.e., $c_j : S \to \mathbb{R}$.

A joint action $s \in S$ is a *pure Nash Equilibrium* (NE) if no player $j \in N$ can benefit from unilaterally deviating from his action to another action, i.e., $\forall j \in N \; \forall a \in S_j \; : \; c_j(s_{-j}, a) \geq c_j(s)$. A pure joint action of a coalition $\Gamma \subseteq N$ specifies an action for each player in the coalition, i.e., $s'_\Gamma \in \times_{j \in \Gamma} S_j$. A joint action $s \in S$ is not resilient to a *pure* deviation of a coalition $\Gamma$ if there is a pure joint action $s'_\Gamma$ of $\Gamma$ such that $c_j(s_{-\Gamma}, s'_\Gamma) < c_j(s)$ for every $j \in \Gamma$ (i.e., the players in the coalition can deviate in such a way that *each* player strictly reduces its cost). In this case we say that the deviation to $s' = (s_{-\Gamma}, s'_\Gamma)$ is a profitable deviation for coalition $\Gamma$.

A pure joint action $s \in S$ is *resilient to pure deviations by coalitions* if there is no coalition $\Gamma \subseteq N$ that has a profitable deviation from $s$.

**Definition 2.1** *A pure* strong equilibrium (SE) *is a pure joint action that is resilient to pure deviations of coalitions.*

Clearly, a strong equilibrium is a refinement of the notion of Nash equilibrium (in particular, if $s$ is a strong equilibrium, it is resilient to deviations of coalitions of size 1, which coincides with the definition of NE).

## 2.2 Job Scheduling on Identical Machines

A job scheduling setting with identical machines is characterized by a set of machines $M = \{M_1, \ldots, M_m\}$, a set $\{1, \ldots, n\}$ of jobs, where a job $j$ has processing time $p_j$. An assignment method produces an assignment $s$ of jobs into machines, where $s_j \in M$ denotes the machine job $j$ is assigned to. The assignment is referred to as a schedule or a configuration (we use the two terms interchangeably). The load of a machine $M_i$ in a schedule $s$ is the sum of the processing times of the jobs assigned to $M_i$, that is $L_i(s) = \sum_{j:s_j=M_i} p_j$. For a set of jobs $\Gamma$, let $s(\Gamma) = \bigcup_{j\in\Gamma} \{s_j\}$ denote the set of machines on which the members of $\Gamma$ are assigned in schedule $s$.

The *makespan* of a schedule is the load on the most loaded machines. A social optimum minimizes the makespan, i.e., $OPT = \min_s \text{makespan}(s)$.

For each job scheduling setting define a *job scheduling game* with the jobs as players. The action space $S_j$ of player $j \in N$ are all the individual machines, i.e., $S_j = M$. The joint action space is $S = \times_{j=1}^{n} S_j$. A joint action $s \in S$ constitutes a schedule. In a schedule $s \in S$ player $j \in N$ selects machine $s_j$ as its action and incurs a cost $c_j(s)$, which is the load on the machine $s_j$, i.e., $c_j(s) = L_i(s)$, where $s_j = M_i$. In a job scheduling game, the makespan is also the highest cost among all players. Formally, $\text{makespan}(s) = \max_j c_j(s)$.

Let $s$ and $s'$ be two configurations. Let $P^{s,s'}_{i_1,i_2}$ be a binary indicator whose value is 1 if there is a job $j$ such that $s_j = M_{i_1}$ and $s'_j = M_{i_2}$ (i.e., if there is a job that chooses $M_{i_1}$ in $s$ but $M_{i_2}$ in $s'$), and 0 otherwise. When clear from the context, we abuse notation and denote $P^{s,s'}_{i_1,i_2}$ by $P_{i_1,i_2}$. In addition, we denote $L_i(s)$ and $L_i(s')$ by $L_i$ and $L'_i$, respectively.

Let $s' = (s_{-\Gamma}, s'_\Gamma)$ be a profitable deviation from $s$ for a coalition $\Gamma$. The *improvement ratio* of a job $j \in \Gamma$ such that $s_j = M_{i_1}$ and $s'_j = M_{i_2}$ (i.e., a job migrating from machine $M_{i_1}$ to machine $M_{i_2}$) is denoted by $IR^{s,s'}(j) = L_{i_1}(s)/L_{i_2}(s')$. Clearly, for any job $j \in \Gamma$, $IR^{s,s'}(j) > 1$. The *damage ratio* of a job $j \notin \Gamma$ such that $s_j = s'_j = M_i$ is denoted by $DR^{s,s'}(j) = L_i(s')/L_i(s)$.

If $s_j \neq s'_j$, we say that job $j$ *migrates* in the deviation. Note that, in our terminology, a job can be a member of a profitable deviation even if it does not migrate in the deviation. Yet, every job that migrates in the deviation is a member of the deviating coalition by definition.

**Definition 2.2** *Given schedules $s$ and $s' = (s_{-\Gamma}, s'_\Gamma)$, the* minimal improvement ratio *of $s'$ with respect to $s$ is $IR_{min}(s, s') = min_{j\in\Gamma} IR^{s,s'}(j)$. In addition, the minimal improvement ratio of a schedule $s$ is $IR_{min}(s) = max_{s'=(s_{-\Gamma},s'_\Gamma),\Gamma\subseteq N} IR_{min}(s, s')$.*

*Given schedules $s$ and $s' = (s_{-\Gamma}, s'_\Gamma)$, the* maximal improvement ratio *of $s'$ with respect to $s$ is $IR_{max}(s, s') = max_{j\in\Gamma} IR^{s,s'}(j)$. In addition, the maximal improvement ratio of a schedule $s$ is $IR_{max}(s) = max_{s'=(s_{-\Gamma},s'_\Gamma),\Gamma\subseteq N} IR_{max}(s, s')$.*

*Given schedules $s$ and $s' = (s_{-\Gamma}, s'_\Gamma)$, the* maximal damage ratio *of $s'$ with respect to $s$ is $DR_{max}(s, s') = max_{j\in N} DR^{s,s'}(j)$. In addition, the maximal damage ratio of a schedule $s$ is $DR_{max}(s) = max_{s'=(s_{-\Gamma},s'_\Gamma),\Gamma\subseteq N} DR_{max}(s, s')$.*

In particular, we can define the notion of $\alpha$-SE (Albers, 2009) in terms of the minimum improvement ratio as follows:

**Definition 2.3** *A schedule $s$ is an* $\alpha$-strong equilibrium ($\alpha$-SE) *if $IR_{min}(s) \leq \alpha$.*

We next provide several useful observations and claims we shall use in the sequel. We refer to a profitable deviation from an NE-schedule as an *NE-originated profitable deviation.* Similarly, a profitable deviations from a schedule produced by the LPT rule is referred to as an *LPT-originated profitable deviation.*

The first observation shows that in any NE-originated profitable deviation, if a job migrates to some machine, some other job must migrate out of that machine. Formally:

**Observation 2.4** *Let $s$ be an NE and let $s' = (s_{-\Gamma}, s'_{\Gamma})$ be a profitable deviation. If $s'_j = M_i$ for some $j \in \Gamma$, then $\exists j' \in \Gamma$ such that $s_{j'} = M_i$ and $s'_{j'} = M_{i'}$ for some $i' \neq i$.*

This is obvious, since if job $j$ strictly decreases its cost by migrating to a machine that no other job leaves, it can also profitably migrates unilaterally, contradicting $s$ being an NE.

We next define a special deviation structure, called a *flower structure* in which all the deviations are from or to the most loaded machine.

**Definition 2.5** *Let $M_1$ be the most loaded machine in a given schedule $s$. We say that a deviation $s'$ obeys the* flower structure *if for all $i > 1$, $P^{s,s'}_{1,i} = P^{s,s'}_{i,1} = 1$ and for all $i, j > 1$, $P^{s,s'}_{i,j} = 0$ (See Figure 2).*

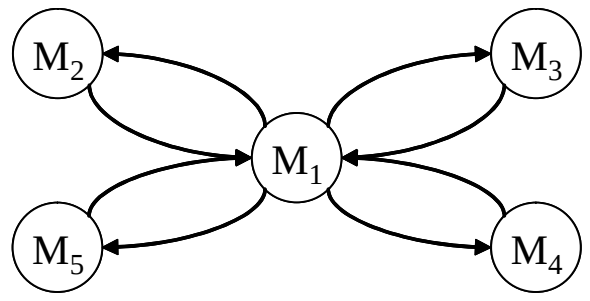

Figure 2: A graph representation of a coalition on 5 machines obeying the flower structure. There is an edge from $M_i$ to $M_j$ if and only if $P^{s,s'}_{i,j} = 1$.

In particular, for $m = 3$, a deviation from $s$ to $s'$ obeys the flower structure if $P^{s,s'}_{1,2} = P^{s,s'}_{2,1} = P^{s,s'}_{1,3} = P^{s,s'}_{3,1} = 1$ and $P^{s,s'}_{2,3} = P^{s,s'}_{3,2} = 0$. Recall that for simplicity of presentation, we write in the sequel $P_{i,j}$ to denote $P^{s,s'}_{i,j}$ and we also write $L_i$ and $L'_i$ to denote $L_i(s)$ and $L_i(s')$, respectively.

**Claim 2.6** *Any NE-originated profitable deviation on three machines obeys the flower structure.*

*Proof*: Let $s$ be an NE and $M_1$ be the most loaded machine in $s$, and let $s'$ be a profitable deviation. We first show that $P_{2,3} = P_{3,2} = 0$. Assume first that both $P_{2,3} = P_{3,2} = 1$. Thus, $L'_2 < L_3$ and $L'_3 < L_2$. Clearly, since the total load does not change, $\sum_i L_i = \sum_i L'_i$. Therefore, it must hold that $L'_1 > L_1$. However, no profitable deviation can increase the load on the most loaded machine. A contradiction. Therefore, at most one of $P_{2,3}, P_{3,2}$ can be 1. Assume w.l.o.g that $P_{2,3} = 1$. By Observation 2.4 some job leaves $M_3$, and by the above it cannot be to $M_2$. Thus, it must be that $P_{3,1} = 1$. Similarly, some job leaves $M_1$. If $P_{1,2} = 1$, then we get that $L'_1 < L_3$, $L'_2 < L_1$, and $L'_3 < L_2$, contradicting $\sum_i L_i = \sum_i L'_i$.

If $P_{1,3} = 1$ then we get that $L'_1 < L_3$, $L'_2 < L_2$ (no job is added to $M_2$), and $L'_3 < L_1$, contradicting $\sum_i L_i = \sum_i L'_i$ again. Thus, $P_{2,3} = 0$. The proof of $P_{3,2} = 0$ is analogous.

It remains to show that $P_{1,2} = P_{1,3} = P_{2,1} = P_{3,1} = 1$. We know that all three machines are assigned jobs in $s'$ that were not assigned to them in $s$. By the above $P_{2,3} = P_{3,2} = 0$. By Observation 2.4 some job leaves each of $M_2, M_3$, therefore, $P_{2,1} = P_{3,1} = 1$. Also, some job leaves $M_1$, thus at least one of $P_{1,2}, P_{1,3}$ equals 1. Assume w.l.o.g that $P_{1,2} = 1$. We show that also $P_{1,3} = 1$. In particular, we show that $L'_3 > L_3$, and since $P_{2,3} = 0$ it must be that $P_{1,3} = 1$. Assume the opposite, that is $L'_3 \leq L_3$. We already know that $P_{1,2} = P_{2,1} = 1$. Thus, $L'_2 < L_1$, $L'_1 < L_2$, and by our assumption $L'_3 \leq L_3$. That is, $\sum_i L'_i < \sum_i L_i$. A contradiction. □

It is known that any NE schedule on two identical machines is also an SE (Andelman et al., 2007). By the above claim, at least four jobs change machines in any profitable deviation on three machines. Clearly, at least four jobs change machines in any coalition on $m > 3$ machines. Therefore,

**Corollary 2.7** *Every NE-schedule is stable against deviations by coalitions of size three or less.*

The next two propositions further characterize any coalition deviation on three machines. We show that while $M_1$ is the most loaded machine before the deviation, it becomes the least loaded after the deviation.

**Proposition 2.8** *In any NE-originated deviation on three machines, the loads on the two less loaded machines in increasing, that is, $L'_2 > L_2$ and $L'_3 > L_3$.*

*Proof*: Assume on the contrary that $L'_2 \leq L_2$. By Claim 2.6, $P_{1,3} = P_{3,1} = 1$. Thus, $L'_3 < L_1$, $L'_1 < L_3$, and by our assumption $L'_2 \leq L_2$. That is, $\sum_i L'_i < \sum_i L_i$. A contradiction. The proof of $L'_3 > L_3$ is analogous. □

**Proposition 2.9** *In any NE-originated deviation on three machines the most loaded machine becomes the least loaded machine, that is, $L'_1 < \min(L'_2, L'_3)$.*

*Proof*: By Claim 2.6, $P_{1,2} = 1$, and thus $L'_1 < L_2$. By the above proposition, $L_2 < L'_2$. Thus, $L'_1 < L'_2$. The proof of $L'_1 < L'_3$ is symmetric. □

## 3. $\alpha$-Strong Equilibrium

In this section, the stability of configurations is measured by the minimal improvement ratio measure. We first provide a complete analysis (i.e. matching upper and lower bounds) for three identical machines[5] for both NE and LPT. For arbitrary $m$, we provide an upper bound for NE and LPT, and show that the lower bounds for $m = 3$ hold for any $m$.

**Theorem 3.1** *Any NE schedule on three machines is a $\frac{5}{4}$-SE.*

5. We note that for unrelated machines, the improvement ratio cannot be bounded within any finite factor even for two machines. This can be seen by a simple example of two jobs and two machines, where the load vector of job 1 is $(1, \varepsilon)$, and the load vector of job 2 is $(\varepsilon, 1)$. If job $i$ is assigned to machine i (for $i = 1, 2$), the resulting configuration is an NE, with load 1 on each machine. However, both jobs can reduce their load from 1 to $\varepsilon$ by swapping.

*Proof*: Let $s$ be an NE-configuration on three machines, and let $r = IR_{min}(s)$. By Claim 2.6, the deviation obeys the flower structure. Therefore: $L'_1 \leq L_2/r$ , $L'_1 \leq L_3/r$ , $L'_2 \leq L_1/r$ , and $L'_3 \leq L_1/r$. Let $P = \sum_j p_j$ (also $= L_1 + L_2 + L_3$). Summing up the above inequalities we get $r \leq (L_1 + P)/(L'_1 + P)$.

**Proposition 3.2** *The load on the most loaded machine is at most half of the total load, that is, $L_1 \leq P/2$.*

*Proof*: Let $g = max(L_1 - L_2, L_1 - L_3)$. By the flower structure, there are at least two jobs on $M_1$, thus $g \leq L_1/2$ - since otherwise some job would benefit from leaving $M_1$, contradicting the NE. By definition of $g$, we know that $2L_1 \leq L_2 + L_3 + 2g$, and since $2g \leq L_1$, we get that $L_1 \leq P/2$. □

Distinguish between two cases:

1. $L'_1 \geq P/5$: in this case $r \leq (L_1 + P)/(L'_1 + P) \leq (3P/2)/(6P/5) = 5/4$.
2. $L'_1 < P/5$: It means that $L'_2 + L'_3 > 4P/5$ ($M_2$ and $M_3$ have the rest of the load), that is, at least one of $L'_2, L'_3 > 2P/5$. W.l.o.g. let it be $M_2$. By the flower structure some job from $M_1$ migrates $M_2$. This job's improvement ratio is $L_1/L'_2$, which, by Proposition 3.2, is less than $(P/2)/(2P/5) = 5/4$. Thus, again, $r < 5/4$.

□

The above analysis is tight as shown in Figure 1. Moreover, this lower bound can be extended to any $m > 3$ by adding $m - 3$ machines and $m - 3$ heavy jobs assigned to these machines. Thus,

**Theorem 3.3** *For $m \geq 3$, there exists an NE schedule $s$ such that $IR_{min}(s) = \frac{5}{4}$.*

For LPT configurations, the bound on the minimum improvement ratio is lower. The proof of the following theorems appear in Appendix A.

**Theorem 3.4** *Any LPT schedule on three machines is a ($\frac{1}{2} + \frac{\sqrt{6}}{4} \approx 1.1123$)-SE.*

**Theorem 3.5** *For any $m \geq 3$, there exists an LPT schedule $s$ such that $IR_{min}(s) = \frac{1}{2} + \frac{\sqrt{6}}{4}$.*

We next provide upper bounds for arbitrary $m$. Our analysis is based on drawing a connection between the makespan approximation and the SE-approximation. Assume that a given schedule is an $r$-approximation for the minimum makespan. We show that under some conditions on the original schedule, if a subset of jobs form a coalition for which $IR_{min} > r$, then, by considering only a subset of machines, it is possible to get a schedule which is apparently better than the optimal one. We first define the set of assignment rules for which the above connection exists.

**Definition 3.6** *Let $s$ be a schedule of an instance $I = \langle N, M \rangle$. Given $\hat{M} \subseteq M$, let $\hat{I} = \langle \hat{N}, \hat{M} \rangle$ be an instance induced by $s, \hat{M}$ such that $\hat{N} = \{j | s_j \in \hat{M}\}$. An assignment method, $A$, is said to be* subset-preserving *if for any $I$ and $\hat{M} \subseteq M$, it holds that $s_j = \hat{s}_j$ for any $j \in \hat{N}$, where $s$ and $\hat{s}$ are the assignments produced by $A$ on the instances $I$ and $\hat{I}$, respectively.*

**Claim 3.7** *LPT is a subset-preserving method.*

*Proof*: The proof is by induction on the number of the jobs in $\hat{N}$. We show that for any $k \neq \hat{N}$, the first $k$ jobs in $N'$ are assigned on the same machine when LPT is executed on input $I$ and on input $\hat{I}$. Note that since $\hat{N}$ is a sublist of $N$, the jobs in $\hat{N}$ are in the same order as in $N$. The first job is scheduled on the first empty machine among $\hat{M}$. For any other job $j \in \hat{N}$, by the induction hypothesis, when $j$ is scheduled, the load on each of the machines is identical to the load of the corresponding machines at the time $j$ was scheduled as a member of $N$. This load is generated only by jobs in $\hat{N}$ that come before $j$ in $N$. Therefore, by LPT, $j$ is scheduled on the least loaded machine among the machines $\hat{M}$, that is, $s_j = \hat{s}_j$ . We assume that LPT uses a deterministic tie-breaking rule if there are several least loaded machines in $\hat{N}$. Therefore $s_j = \hat{s}_j$ also in this case. □

**Lemma 3.8** *Let A be an assignment method that is (i) subset-preserving, (ii) yields Nash equilibrium, and (iii) approximates the minimum makespan within a factor of r, where $r \geq 1$ is non-decreasing in m. Then, A produces an r-SE.*

*Proof*: Assume for contradiction that there exists an instance $I$ such that in the schedule $s$ produced by $A$, there exists a coalition in which the improvement ratio of every member is greater than $r$. Let $\Gamma$ be such a coalition of minimum size. If there is a job $j \in \Gamma$ that does not migrate, then the set of jobs $\Gamma \setminus \{j\}$ is a smaller coalition, contradicting the minimality of $\Gamma$; therefore, for every $j \in \Gamma$, it holds that $s_j \neq s'_j$. We next show that $s(\Gamma) = s'(\Gamma)$. First, $s(\Gamma) \subseteq s'(\Gamma)$, that is, for every machine in $s(\Gamma)$ from which a job $j$ migrates, there must exist a job migrating to it, otherwise, $\Gamma \setminus \{j\}$ is a smaller such coalition, contradicting the minimality of $\Gamma$. Second, $s'(\Gamma) \subseteq s(\Gamma)$, that is, for every machine to which a job $j$ migrates, there must exist a job migrating from it (otherwise job $j$ can improve unilaterally, in contradiction to $s$ being an NE). Given that $s(\Gamma) = s'(\Gamma)$, denote this set of machines by $\hat{M}$, and let $\hat{m} = |\hat{M}|$. Finally, let $\hat{N} \subseteq N$ be the set jobs assigned to machines in $\hat{M}$ by $A$,

Consider the instance $\hat{I} = \langle \hat{N}, \hat{M} \rangle$. Since $A$ is subset-preserving, the jobs of $\hat{N}$ are scheduled by $A$ on $\hat{M}$ in $\hat{I}$ exactly as their schedule on $\hat{M}$ when scheduled as part of $I$. In particular, when $\hat{I}$ is scheduled by $A$, the same deviation of $\Gamma$ exists, in which every job in $\Gamma$ improves by a factor greater than $r$, and all the machines in $\hat{M}$ take part in it. In other words, for any pair of machines $i, j$, such that $P_{i,j} = 1$, we have $L_i/L'_j > r(m) \geq r(\hat{m})$, where $r(m)$ is the approximation ratio of $A$ on $m$ machines. On the other hand, since $A$ produces an $r(\hat{m})$-approximation, for any machine $i$, $L_i \leq r(\hat{m})OPT(\hat{I})$, where $OPT(\hat{I})$ is the minimum possible makespan of $\hat{I}$ on $\hat{M}$ machines. Therefore, if $P_{i,j} = 1$ then $r(\hat{m}) < L_i/L'_j \leq \frac{r(\hat{m})OPT(\hat{I})}{L'_j}$. This implies that for any machine $j$ that receives at least one job, $L'_j < OPT(\hat{I})$.

However, since at least one job has migrated to each of the $\hat{m}$ participating machines, after the deviation the machines $\hat{M}$ are assigned all the jobs of $\hat{N}$ and they all have load less than $OPT(\hat{I})$. A contradiction. □

Let $s$ be an NE on $M$ machines. Clearly, for any $\hat{M} \subseteq M$, the induced schedule of $s$ on the set of machines $\hat{M}$ is also an NE. Also, it is known that any NE provides a $(2 - \frac{2}{m+1})$-approximation to the makespan (Finn & Horowitz, 1979; Schuurman & Vredeveld, 2007). This implies that Lemma 3.8 can be applied with $r = 2 - \frac{2}{m+1}$ to any assignment that yields an NE. Therefore,

**Corollary 3.9** *Any NE schedule on $m$ identical machines is a $(2 - \frac{2}{m+1}) - SE$.*

The next result is a direct corollary of Lemma 3.8, Claim 3.7, and the fact that LPT provides a $(\frac{4}{3} - \frac{1}{3m})$-approximation to the makespan (Graham, 1969).

**Corollary 3.10** *Any schedule produced by LPT on $m$ identical machines is a $(\frac{4}{3} - \frac{1}{3m}) - SE$.*

The above bounds are not tight, but the gap between the lower and upper bounds is only a small constant.

## 4. Maximum Improvement Ratio

In this section, we analyze the maximum improvement ratio measure. We provide a complete analysis for NE configurations and any $m \geq 3$, and for LPT configurations on three machines. The lower bound for LPT on three machines can be extended to arbitrary $m$. In contrast to the other measures we consider in this paper, where NE and LPT differ only by a small constant, it turns out that with respect to the maximum improvement ratio, NE and LPT are significantly different. While the improvement ratio of NE-originated deviations can be arbitrarily high, for deviations from LPT configurations, the highest possible improvement ratio of any participating job is less than $\frac{5}{3}$.

**Theorem 4.1** *Fix $r \geq 1$. For any $m \geq 3$ machines, there exists an instance with $m$ machines and an NE $s$ such that $IR_{max}(s) > r$.*

*Proof*: Given $r$, consider the NE-configuration on three machines given in Figure 3(a). The coalition consists of $\{1, 1, 2r, 2r\}$. Their improved schedule is given in Figure 3(b). The improvement ratio of the jobs of load 1 is $2r/2 = r$. For $m > 3$, dummy machines and jobs can be added. □

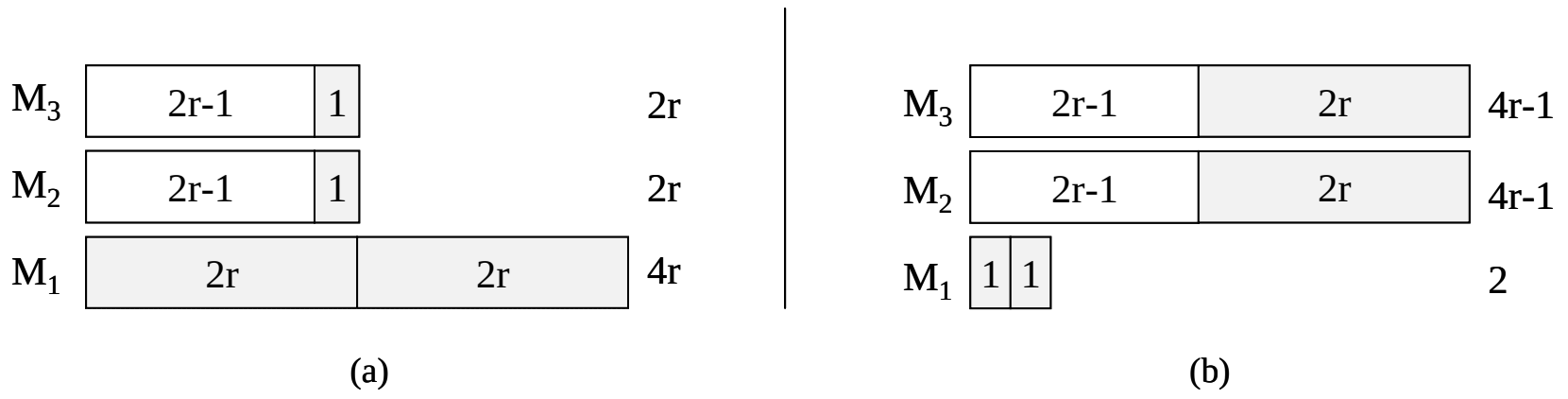

Figure 3: An NE-originated deviation in which the jobs of load 1 have improvement ratio $r$.

In contrast to NE-originated deviations, for LPT-originated deviations we are able to bound the maximum improvement ratio by a small constant. The proof of the following claim is given in Appendix A.

**Theorem 4.2** *Let $s$ be an LPT schedule on three machines. It holds that $IR_{max}(s) \leq \frac{5}{3}$.*

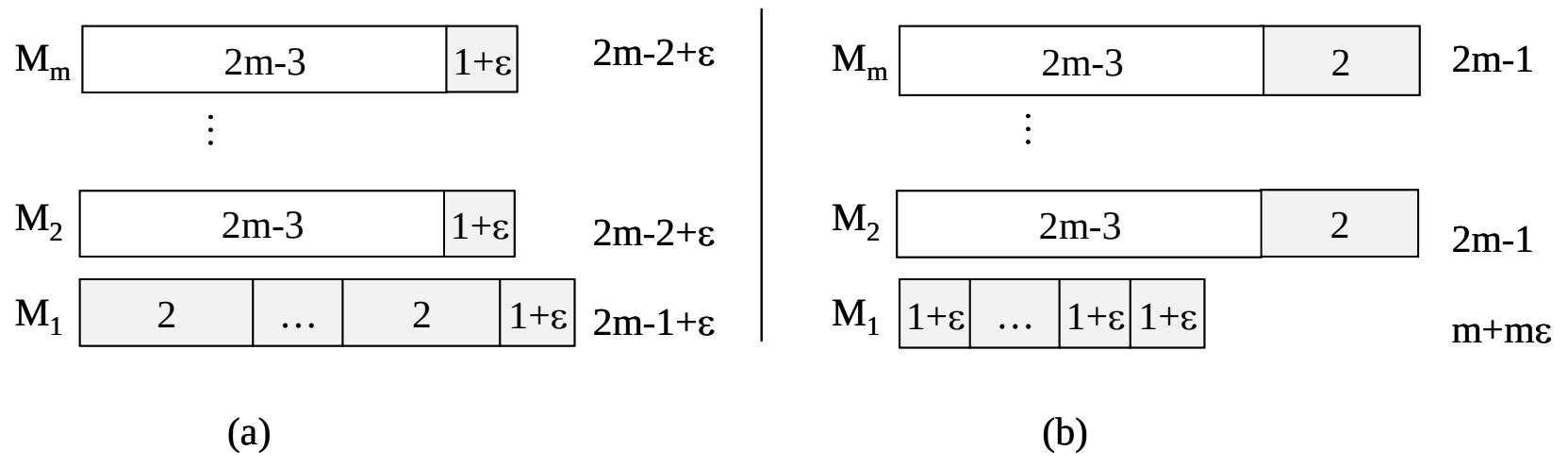

Figure 4: An LPT-originated deviation on $m$ machines in which the job of load $1+\varepsilon$ assigned to $M_1$ has improvement ratio arbitrarily close to $2 - \frac{1}{m}$.

The above bound is tight, as demonstrated in Figure 4 for $m = 3$ (where the improvement ratio is $2 - \frac{1}{m} = \frac{5}{3}$). Moreover, this figure shows that this lower bound can be generalized for any $m \geq 3$. Note that the coalitional deviation in this example obeys the flower structure. We believe that this example is tight, as the flower structure seems to enable the largest possible decrease in the load of a single machine. The job of load $1+\varepsilon$ that remains on $M_1$ improves its cost from $2m-1+\varepsilon$ to $m(1+\varepsilon)$, that is, for this job, $j$, $IR(j) = \frac{2m-1+\varepsilon}{m(1+\varepsilon)} = 2 - \frac{1}{m} - \delta$. Formally,

**Theorem 4.3** *For any $m \geq 3$, there exists an LPT configuration $s$ such that $IR_{max}(s) = 2 - \frac{1}{m} - \delta$ for an arbitrarily small $\delta > 0$.*

## 5. Maximum Damage Ratio

In this section, we provide results for the maximum damage a profitable deviation can impose on jobs that do not take part in the coalition. Formally, the quality of a configuration is measured by $max_{j \notin \Gamma} DR(j)$. We provide a complete analysis for NE and LPT configurations and any $m \geq 3$. Once again, we find that LPT provides a strictly better performance guarantee compared to general NE: the cost of any job in an LPT schedule cannot increase by a factor $\frac{3}{2}$ or larger, while it can increase by a factor arbitrarily close to 2 for NE schedules.

**Theorem 5.1** *For any $m$, $DR_{max}(s) < 2$ for every NE configuration $s$.*

*Proof*: Let $s' = (s'_\Gamma, s_{-\Gamma})$ be a profitable deviation, and let $M_1$ be the most loaded machine among the machines that either a job migrated from or a job migrated into. By Observation 2.4, there must be a job that migrated out of $M_1$. This implies that there must be at least two jobs on $M_1$ in $s$, since if there were a single job, it could not benefit from any deviation. Therefore, there exists a job $j$ such that $s_j = M_1$ and $p_j \leq L_1/2$. Using the fact that $s$ is an NE once again, we get that for any machine $i \neq 1$, $L_i \geq L_1/2$ (otherwise job $j$ can improve by unilaterally migrating to $M_i$).

In addition, for every machine $i$ to which a job migrates, it must hold that $L'_i < L_1$. This is because a job that migrated to $M_i$ left some machine $j$ with load $L_j \leq L_1$. Combining the above bounds, we get that for every job $j$ that stays on some machine $i$ to which a job migrates it holds that $DR^{s,s'}(j) = L'_i/L_i < L_1/L_i \leq (2L_i)/L_i = 2$. □

The above analysis is tight as shown in Figure 3: The damage ratio of the jobs of load $2r-1$ is $(4r-1)/(2r)$, which can be arbitrarily close to 2. Formally,

**Theorem 5.2** *For any $m \geq 3$, there exists an NE configuration $s$ such that $DR_{max}(s) = 2 - \delta$ for an arbitrarily small $\delta > 0$.*

For LPT configurations we obtain a smaller bound:

**Theorem 5.3** *For any $m$, $DR_{max}(s) < \frac{3}{2}$ for every LPT configuration $s$.*

*Proof*: Let $s' = (s'_\Gamma, s_{-\Gamma})$ be a profitable deviation, and let $M_1$ be the most loaded machine among the machines that either a job migrated from or a job migrated into. Since every LPT configuration is an NE, $M_1$ must have at least two jobs (following the same arguments as in the proof of Theorem 5.1). Assume w.l.o.g that the lightest (also last) job assigned to $M_1$ has load 1, and denote this job the "1-job". This assumption is valid because the minimum improvement ratio is invariant to linear transformations. Let $\ell = L_1 - 1$. Since $s$ is an LPT configuration, for every machine $i$, it must hold that $L_i \geq \ell$ (otherwise, the 1-job would have been assigned to a different machine). In addition, since for every machine $j$ from which a job migrates, $L_j \leq L_1$, it must hold that for every machine $i$ to which a job migrates $L'_i < \ell + 1$. We distinguish between two cases.

case (a): $\ell \geq 2$. Then, for every machine $M_i$ to which a job migrates, $L'_i/L_i < \frac{\ell+1}{\ell} \leq \frac{3}{2}$.

case (b): $\ell < 2$. We show that no profitable deviation exists in this case. If $\ell < 2$, then $M_1$ has exactly 2 jobs, of loads $\ell$ and 1, since LPT assigns the jobs in a non-increasing order. By LPT, every other machine must have (i) one job of load at least $\ell$ (and possibly other small jobs), or (ii) two jobs of load at least 1 (and possible other small jobs). Let $k$ and $k'$ be the number of machines of type (i) and (ii), respectively (excluding $M_1$). Thus, there is a total of $k+1$ jobs of load $\ell$ and $2k'+1$ jobs of load 1. After the deviation, no machine can have jobs of load $\ell$ and 1 together, nor can it have three jobs of load 1. The $k+1$ machines assigned with the $k+1$ jobs of load $\ell$ after the deviation cannot be assigned any other job of load $x$. So, we end up with $2k'+1$ jobs of load 1 that should be assigned to $k'$ machines. Thus, there must be a machine with at least three jobs of load 1. Contradiction. □

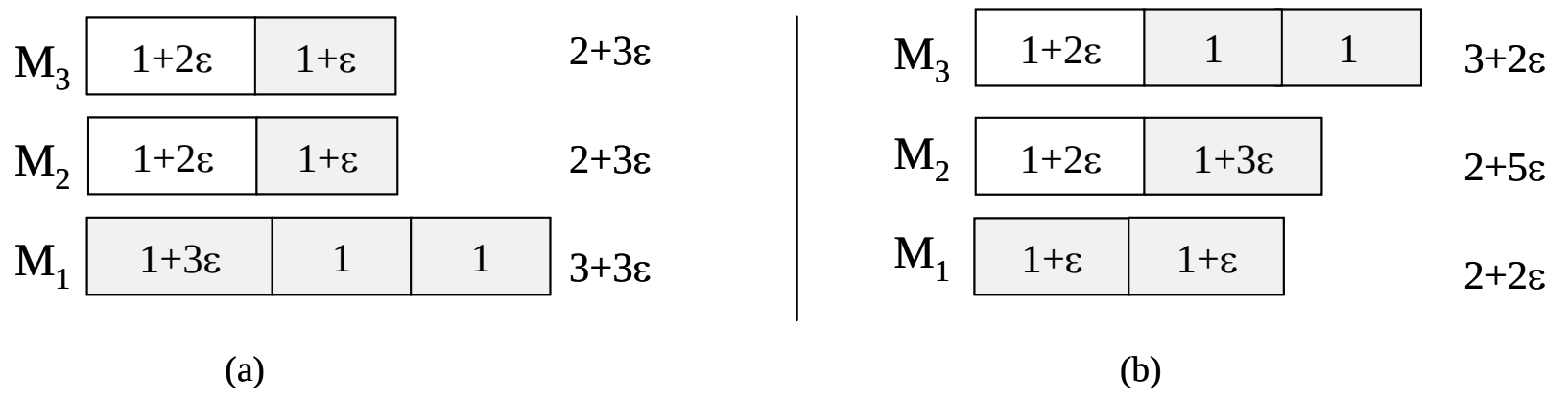

Figure 5: An LPT-originated deviation, in which the damage ratio of the job of load $1+2\varepsilon$ on $M_3$ is arbitrarily close to $\frac{3}{2}$.

The above analysis is tight as shown in Figure 5. Moreover, by adding dummy machines and jobs it can be extended to any $m \geq 3$. Formally,

**Theorem 5.4** *For any $m \geq 3$, there exists an LPT configuration $s$ such that $DR_{max}(s) = \frac{3}{2} - \delta$ for an arbitrarily small $\delta > 0$.*

## 6. Approximation Scheme

In this section we present a polynomial time approximation scheme that provides a $(1+\varepsilon)$-SE. The PTAS can be applied to any fixed number of machines.

**Definition 6.1** *A vector $(l_1, l_2, \dots l_m)$ is smaller than $(\hat{l}_1, \hat{l}_2, \dots \hat{l}_m)$* lexicographically if for some $i$, $l_i < \hat{l}_i$ and for all $b < i$, $l_b = \hat{l}_b$ . A configuration $s$ is lexicographically smaller than $\hat{s}$ if the vector of machine loads $L(s) = (L_1(s), \dots, L_m(s))$, sorted in non increasing order, is smaller lexicographically than $L(\hat{s})$, sorted in non increasing order.

The PTAS combines a lexicographically minimal assignment of the longest $k$ jobs with the LPT rule applied to the remaining jobs. The value of $k$ depends on the desired approximation ratio (to be defined later).

Formally, the algorithm $A_k$ is defined as follows:

1. Find a lexicographically minimal assignment of the longest $k$ jobs.
2. Add the remaining jobs greedily using the LPT rule.

In particular, since a lexicographically minimal assignment minimizes the makespan (given by the load on the most loaded machine), the above algorithm is a PTAS for the minimum makespan problem, as it is a restriction of the known PTAS of Graham (1966). In Graham's algorithm, in step 1, the first $k$ jobs are scheduled in a way that minimizes the makespan. In our scheme, the requirement on the schedule of the long jobs is more strict. In particular, as shown by Andelman et al. (2007), the schedule of the longest $k$ jobs is an SE of this sub-instance.

Given $\varepsilon$, let $A_\varepsilon$ denote the above algorithm with $k = \lceil \frac{m}{\varepsilon} \rceil$. We first show that for any subset of machines $\hat{M} \subseteq M$, $A_\varepsilon$ provides a $(1+\varepsilon)$-approximation to the makespan of the subset of jobs scheduled on $\hat{M}$ . Formally,

**Lemma 6.2** *Let $I = \langle N, M \rangle$ be an instance of job scheduling with machines $M$ and jobs $N$. Let $s$ be an output of $A_\varepsilon$ on $I$. For a given $\hat{M} \subseteq M$, let $\hat{N} \subseteq N$ be the set of jobs scheduled on $\hat{M}$ in $s$. Consider the instance $\hat{I} = \langle \hat{N}, \hat{M} \rangle$. Let $\hat{s}$ be the assignment of $\hat{I}$ induced by $s$. Then $\hat{s}$ is $(1+\varepsilon)$-approximation for the makespan of $\hat{I}$.*

*Proof*: Let $L^{A_\varepsilon}_{max}(M)$ denote the largest completion time of a machine in a set $M$ in the schedule produced by $A_\varepsilon$, and let $OPT(I)$ denote the minimum makespan of $I$. Let $\hat{T}$ denote the largest completion time of a long job in $\hat{N}$ scheduled on $\hat{M}$ in the minimal lexicographic schedule found in step 1. Since $s$ is a minimal lexicographic assignment, $\hat{T}$ is the minimum makespan of the long jobs of $\hat{N}$. In particular, $\hat{T}$ is a lower bound for $OPT(\hat{I})$, thus, if the makespan on $\hat{M}$ is not increased in the second step, that is, $L^{A_\varepsilon}_{max}(\hat{M}) = \hat{T}$, then $A_\varepsilon$ is optimal for $\hat{I}$. Otherwise, the makespan of $\hat{M}$ is larger than $\hat{T}$. Let $j$ be the job determining the makespan of $\hat{M}$ (the job who completes last in $\hat{N}$). By definition of LPT, this implies that all the machines $\hat{M}$ were busy when job $j$ started its execution (otherwise job $j$ could start earlier). Since the optimal schedule from step 1 has no intended idles, it holds that all the machines $\hat{M}$ are busy during the time interval $[0; L^{A_\varepsilon}_{max}(\hat{M}) - p_j]$

Let $\hat{P} = \sum_{j=1}^{\hat{n}} p_j$ be the total processing time of the $\hat{n}$ jobs in $\hat{N}$. By the above, $\hat{P} \geq \hat{m}(L^{A_\varepsilon}_{max}(\hat{M}) - p_j)$. Also, since the jobs are sorted in non-increasing order of processing times, we have that $p_j \leq p_{k+1}$ and therefore $\hat{P} \geq \hat{m}(L^{A_\varepsilon}_{max}(\hat{M}) - p_{k+1})$. A lower bound for the optimal solution of $\hat{I}$ is a schedule in which the load on the $\hat{m}$ machines is balanced; thus $OPT(\hat{I}) \geq \hat{P}/\hat{m}$, which implies that $L^{A_\varepsilon}_{max}(\hat{M}) \leq OPT(\hat{I}) + p_{k+1}$.

In order to bound $L^{A_\varepsilon}_{max}(\hat{M})$ in terms of $OPT(\hat{I})$, we need to bound $p_{k+1}$ in terms of $OPT(\hat{I})$. We first bound the gap between $OPT(I)$ and $OPT(\hat{I})$. The following assumption is used.

**Claim 6.3** *Let $z$ be the job determining the makespan of $A_\varepsilon(I)$. W.l.o.g., $z$ is not one of the $k$ long jobs.*

*Proof*: Assume that the makespan of $A_\varepsilon(I)$ is determined by one of the long jobs. Let $M_1$ be the machine on which $z$ is scheduled. In particular, $M_1$ processes only long jobs. Fix the schedule on $M_1$ and repeat the PTAS for the remaining jobs and machines with the same value of $k$. Repeat if necessary until the makespan is determined by a job assigned using the LPT rule.

Note that the above algorithm is still polynomial, as the PTAS might be repeated at most $m-1$ times, which is a constant. The approximation ratio is improving for any sub-instance: same number of jobs are considered long, but among a set of fewer jobs, that is, a larger portion of the jobs is scheduled optimally, therefore the approximation ratio proof is valid for the sub-instance. Finally, by merging the last PTAS result with the schedule on the machines holding long jobs only, we get a PTAS for the whole instance, since the long jobs were scheduled optimally in each step. Moreover, the load on each such machine is a lower bound on the makespan of the sub-instance that was considered when the machine gets these jobs. □

**Claim 6.4** $OPT(I) \leq OPT(\hat{I}) + p_{k+1}$.

*Proof*: Let $z$ be the job determining the makespan of $A_\varepsilon(I)$. By Claim 6.3, $z$ can be assumed to be assigned in step 2 (by LPT rule). If $z \in \hat{N}$ then $A_\varepsilon(I) = L^{A_\varepsilon}_{max}(\hat{M})$. Else, the load on any machine in $\hat{M}$ is at least $A_\varepsilon(I) - p_z$, since otherwise job $z$ should have been assigned to one of the machines in $\hat{M}$. Therefore, even if the total load of $\hat{N}$ is balanced among $\hat{M}$, we have that $OPT(\hat{I}) \geq A_\varepsilon(I) - p_z$. Since $p_z \leq p_{k+1}$, and $OPT(I) \leq A_\varepsilon(I)$, we get $OPT(I) \leq A_\varepsilon(I) \leq OPT(\hat{I}) + p_{k+1}$. □

**Claim 6.5** $p_{k+1} \leq OPT(\hat{I}) \left\lceil \frac{m}{k} \right\rceil$.

*Proof*: Consider the $k+1$ longest jobs. In an optimal schedule, some machine is assigned at least $\lceil (k+1)/m \rceil \geq 1 + \lfloor k/m \rfloor$ of these jobs. Since each of these jobs has processing time at least $p_{k+1}$, we conclude that $OPT(I) \geq (1 + \lfloor k/m \rfloor)p_{k+1}$, which implies that $p_{k+1} \leq OPT(I)/(1+\lfloor k/m \rfloor)$. By Claim 6.4, $p_{k+1} \leq OPT(I)/(1+\lfloor k/m \rfloor) \leq (OPT(I')+p_{k+1})/(1+\lfloor k/m \rfloor)$. It follows that $p_{k+1} \leq OPT(\hat{I})\lceil \frac{m}{k} \rceil$. □

Back to the bound on $L^{A_\varepsilon}_{max}(\hat{I})$, we can now conclude that $L^{A_\varepsilon}_{max}(\hat{I}) \leq OPT(\hat{I}) + p_{k+1} \leq OPT(\hat{I})(1+\lceil \frac{m}{k} \rceil) = OPT(\hat{I})(1+\varepsilon)$. □

We can now prove the main result of this section, showing that the schedule $s$ produced by $A_\varepsilon$, is a $(1+\varepsilon)$-SE. The stability is proved in the following theorem. As for the running time, for fixed $m, k$, a minimal lexicographic schedule of the first $k$ jobs can be found in $O(m^k)$ steps. Applying the LPT rule takes additional $O(nlogn)$. For $A_\varepsilon$, we get that the running time of the scheme is $O(m^{m/\varepsilon})$, that is, exponential in $m$ (that is assumed to be constant) and $1/\varepsilon$.

**Theorem 6.6** *$A_\varepsilon$ produces an $(1+\varepsilon)$-SE.*

*Proof*: The proof is similar to the proof of Lemma 3.8. Assume for contradiction that there exists an instance $I$ for $A_\varepsilon$ on $m$ machines, such that in the schedule of $I$ produced by $A_\varepsilon$, there exists a coalition in which the improvement ratio of every member is larger than $1+\varepsilon$. Let $\Gamma$ be such a coalition of minimum size. For every machine from which a job $j$ migrates, there must exists a job migrating to it, otherwise, $\Gamma \setminus \{j\}$ is also a coalition having $IR_{min} > 1+\varepsilon$, in contradiction to the minimality of $\Gamma$. Let $\hat{M}$ denote the set of machines that are part of the coalition, let $\hat{N} \subseteq N$ be the set jobs assigned to $\hat{M}$ by $A_\varepsilon$, and let $\hat{m} = |\hat{M}|$. Consider the instance $\hat{I} = \langle \hat{N}, \hat{M} \rangle$, and the schedule $\hat{s} \subseteq s$. By Lemma 6.2, $\hat{s}$ is a $(1+\varepsilon)$-approximation to the makespan of $\hat{I}$. The coalition $\Gamma$ exists in $\hat{s}$, and all the machines $\hat{M}$ take part in it. Moreover, each of the jobs in $\Gamma$ improves by a factor of more than $(1+\varepsilon)$. In other words, for any pair of machines $i, j$, such that $P_{i,j} = 1$, we have $L_i/L'_j > 1+\varepsilon$. On the other hand, since $\hat{s}$ is a $(1+\varepsilon)$-approximation, for any machine $i$, $L_i \leq (1+\varepsilon)OPT(\hat{I})$. Therefore, if $P_{i,j} = 1$ then $1+\varepsilon < \frac{L_i}{L'_j} \leq \frac{(1+\varepsilon)OPT(\hat{I})}{L'_j}$. In other words, for any machine $j$ that receives at least one job, $L'_j < OPT(\hat{I})$.

However, since at least one job has migrated to each of the $\hat{m}$ participating machines, after the deviation the machines $\hat{M}$ are assigned all the jobs of $\hat{N}$ and they all have load less than $OPT(\hat{I})$. A contradiction. □

We note that for any $\varepsilon \geq 0$, the schedule produced by algorithm $A_\varepsilon$ is an NE. Similar to the stability proof of LPT (Fotakis et al., 2002), it is easy to verify that if some job can benefit from leaving some machine $M_i$ then also the shortest job on this machine can benefit from the same migration. However, independent of whether this short job, of length $p_j$, is assigned in step 1 of the algorithm (as part of a minimal lexicographical schedule of the long job) or in step 2 (by LPT), the gap between $L_i$ and the load on any other machine is at most $p_j$.

## 7. Computational Complexity

It is easy to see that one can determine in polynomial time whether a given configuration is an NE. Yet, for SE, this task is more involved. In this section, we provide some hardness results about coalitional deviations.

**Theorem 7.1** *Given an NE schedule on $m \geq 3$ identical machines, it is NP-hard to determine if it is an SE.*

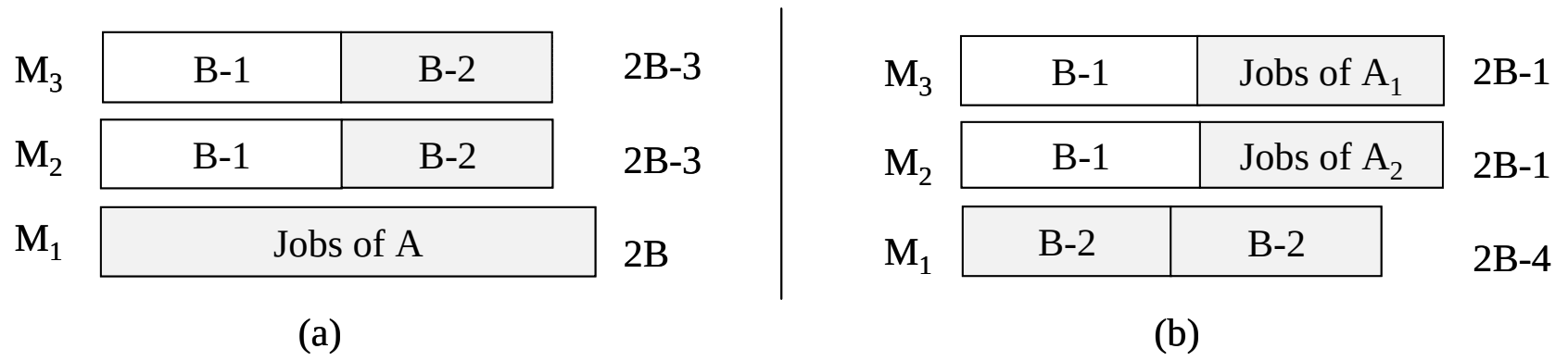

Figure 6: Partition induces a coalition in a schedule on identical machines.

*Proof*: We give a reduction from *Partition*. Given a set $A$ of $n$ integers $a_1, \dots, a_n$ with total size $2B$, and the question whether there is a subset of total size $B$, construct the schedule in Figure 6(a). In this schedule on three machines there are $n+4$ jobs of loads $a_1, \dots, a_n, B-2, B-2, B-1, B-1$. We assume w.l.o.g. that $min_i a_i \geq 3$, else the whole instance can be scaled. Thus, schedule 6(a) is an NE. For $m \geq 3$, add $m-3$ machines each with a single job of load $2B$.

**Claim 7.2** *The NE schedule in Figure 6(a) is an SE if and only if there is no partition.*

*Proof*: If there is a partition into $K_1, K_2$, each having total size $B$, then the schedule in Figure 6(b) is better for the jobs originated from the partition instance and for the two $(B-2)$-jobs. All the partition jobs improved from cost $2B$ to cost $2B-1$, and the $(B-2)$-jobs improved from $2B-3$ to $2B-4$.

Next, we show that if there is no partition then the initial schedule is an SE. By Theorem 2.7, in any action of a coalition on three machines, jobs must migrate to $M_1$ from both $M_2$ and $M_3$. In order to decrease the load from $2B-3$, the set of jobs migrating to $M_1$ must be the set of two jobs of load $B-2$. Also, it must be that all the partition jobs move away from $M_1$ - otherwise, the total load on $M_1$ will be at least $2B-4+3 = 2B-1$, which is not an improvement for the $(B-2)$-jobs. This implies that the jobs of $M_1$ split between $M_2$ and $M_3$. However, since there is no partition, one of the two subsets is of total load at least $B+1$. These jobs will join a job of load $B-1$ to get a total load of at least $2B$, which is not an improvement over the $2B$-load in the initial schedule. □

This establishes the proof of the Theorem. □

A direct corollary of the above proof is the following:

**Corollary 7.3** *Given an NE schedule and a coalition, it is NP-hard to determine whether the coalition can deviate.*

Theorem 7.1 holds for any $m \geq 3$ identical machines. For $m \leq 2$, a configuration is an NE if and only if it is an SE (Andelman et al., 2007), and therefore it is possible to determine whether a given configuration is SE in polynomial time. Yet, the following theorem shows that for the case of unrelated machines, the problem is NP-hard already for $m = 2$. In the unrelated machines environment, the processing time of a job depends on the machine on which it is assigned. For every job $j$ and machine $i$, $p_{i,j}$ denotes the processing time of job $j$ if processed by machine $i$.

**Theorem 7.4** *Given an NE schedule on $m \geq 2$ unrelated machines, it is NP-hard to determine if it is an SE.*

*Proof*: We give a reduction from *Partition*. Given $n$ integers $a_1, \ldots, a_n$ with total size $2B$, and the question whether there is a subset of total size $B$, construct the following instance for scheduling: there are 2 machines and $n+1$ jobs with the following loads (for $\varepsilon < 1/(n-1)$):

$$p_{i,1} = a_i + \varepsilon \text{ and } p_{i,2} = 2a_i + \varepsilon, \forall i \in \{1, \ldots, n\} \ ; \ p_{n+1,1} = B, \text{ and } p_{n+1,2} = 2B + n\varepsilon.$$

Consider the schedule in which all the jobs $1, \ldots, n$ are on $M_1$, and job $n+1$ is on $M_2$. The completion times of both machines are are $2B + n\varepsilon$. It is an NE.

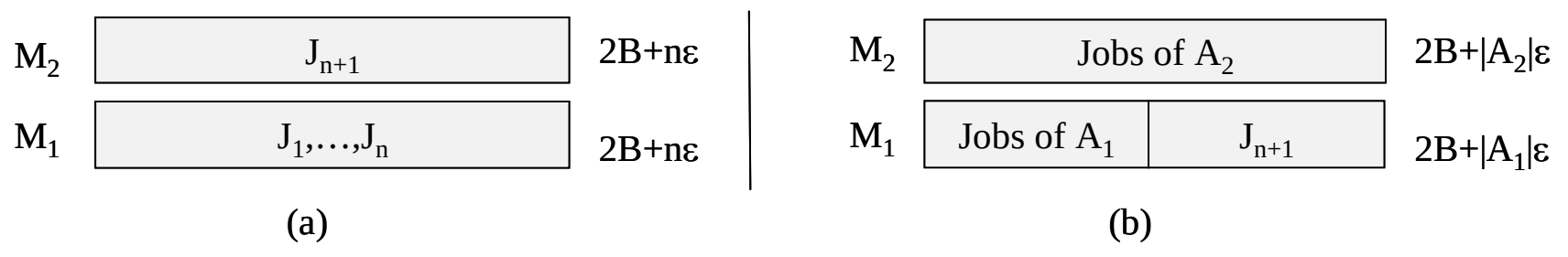

Figure 7: Partition induces a coalition in a schedule on related machines.

**Claim 7.5** *The NE schedule in Figure 7(a) is an SE if and only if there is no partition.*

*Proof*: If there is a partition into $A_1, A_2$, each having total size $B$, then the schedule given in Figure 7(b) is better for everyone. The completion time of $M_1$ is $2B + |A_1|\varepsilon < 2B + n\varepsilon$ and the completion time of $M_2$ is $2B + |A_2|\varepsilon < 2B + n\varepsilon$.

Next, we show that if there is no partition then the initial schedule is an SE. Since there is no partition, in any partition into $A_1, A_2$, one of the two subsets, w.l.o.g., $A_1$ has total size at least $B+1$. $A_1$ will only increase its load by migrating to $M_2$ even alone (bearing a load of at least $2B + 2 + |A_1|\varepsilon$ instead of $2B + n\varepsilon$). Therefore, $A_1$ will not leave $M_1$. However, if $A_1$ stays at $M_1$, job $n+1$ is better-off staying at $M_2$ (since if it migrates, it bears a load of at least $2B + 1 + |A_1|\varepsilon$ which is not smaller than $2B + n\varepsilon$ for any $|A_1|$ and $\varepsilon \leq 1/(n-1)$). □

This establishes the proof of the Theorem. □

A direct corollary of the above proof is the following:

**Corollary 7.6** *Given an NE schedule on unrelated machines and a coalition, it is NP-hard to determine whether the coalition can deviate.*

## 8. Conclusions and Open Problems

In this paper we study how well NE schedules and a special subset of them – those obtained as an outcome of the LPT assignment rule – approximate SE in job scheduling games. We do so using the two measures $IR_{min}$ and $IR_{max}$. In addition, we use the $DR_{max}$ measure to study how hurtful coalitional deviations can be to agents outside the coalition. We present upper and lower bounds for NE and LPT-based schedules, and demonstrate that LPT-based schedules perform better than general NE schedules, where the gap is more significant under the $IR_{max}$ measure. For both NE and LPT, $IR_{min}$ is bounded by a small constant, implying some notion of stability against coalitional deviations (assuming the existence of a transition cost). As for $IR_{max}$, it is bounded by a constant for LPT schedules, but there is no universal bound for NE schedules. Yet, LPT is not the best possible approximation to SE, as demonstrated by the SE-PTAS we design, which computes a schedule with $IR_{min}$ arbitrarily close to 1.

Some of the problems that remain open are:
**1.** For the $IR_{min}$ measure, there is a gap between the upper and lower bounds for $m > 4$ [6].
**2.** For $IR_{max}$ of LPT-originated deviations, and $m \geq 3$ we presented a lower bound of $2 - \frac{1}{m}$ and a matching upper bound of $\frac{5}{3}$ for $m = 3$. Closing this gap for a general $m$ is left as an open problem.
**3.** This paper focuses on the case of identical machines. It would be interesting to study the topic of approximate strong equilibrium in additional job scheduling settings. In particular, the setting of uniformly related machines is part of our ongoing research, where already the case of two machines seems rather involved. Note that, as mentioned in Section 7, for unrelated machines, $IR_{min}$ is unbounded already for two machines.
**4.** Our measures are defined with respect to the strong equilibrium solution concept, where a profitable deviation is defined as one in which every member of the coalition strictly benefits. It would be interesting to consider the measures we introduce here with respect to additional solution concepts, such as *coalition-proof Nash equilibrium* (Bernheim et al., 1987) (which is stable against profitable deviations that are themselves stable against further deviations of sub-coalitions), and also with respect to profitable deviations in which none of the coalition members is worse-off and at least one member is strictly better-off.

In summary, we introduced three general measures for the stability and performance of schedules under coalitional deviations. We believe that these measures can be used to measure the stability and performance of various algorithms to coalitional deviations and their performance in additional settings and games. We hope to see more work that makes use of these measures within the framework of algorithmic game theory. It would be interesting to study in what families of games Nash equilibria approximate strong equilibria as defined by the measures introduced here.

**Acknowledgments.** We thank Leah Epstein and Alon Rosen for helpful discussions. We also thank the anonymous reviewers for their insightful remarks and suggestions. The work was partially supported by the Israel Science Foundation (grant number 1219/09).

## Appendix A. Bounding $IR_{min}$ and $IR_{max}$ in LPT-originated Deviations

We first provide several observation that are valid for any LPT-originated deviation. This observations will be used later in the analysis. Moreover, the observations characterize the coalitions that might exist in schedules produced by the LPT-rule. Combined with the flower structure (that characterizes all NE-originated deviations on three machines), we get that the set of LPT-originated deviation are very limited and must follow a very strict structure.

Let $M_1$ be the most loaded machine. Assume w.l.o.g that the lightest (also last) job assigned to $M_1$ has load 1, and denote this job the "1-job". This assumption is valid because

6. Our paper provides tight bounds for $m = 3$ and the case of $m = 4$ is considered by Chen (2009).

the minimum improvement ratio is invariant to linear transformations. For $i = 2, 3$, denote by $K_i$ the set (and also the total load) of jobs that remain on $M_i$. Denote by $H_{i,j}$ the set (and also total load) of jobs migrating from $M_i$ to $M_j$. For $i = 1$, we let $K_1, H_{1,2}, H_{1,3}$ be as above, but excluding the 1-job.

The next propositions show that the total size of jobs migrating from $M_2, M_3$ to $M_1$ and remaining on $M_2, M_3$ is at least as large as the last job on $M_1$.

**Proposition A.1** *Each of* $H_{2,1}, H_{3,1}$ *is at least* 1.

*Proof*: We show that $H_{2,1} \geq 1$, the proof for $H_{3,1}$ is analogous. Assume for contradiction that $H_{2,1} < 1$. Since LPT schedule the jobs in non-increasing order, all jobs composing $H_{2,1}$ were assigned after the 1-job. Therefore, when the 1-job is assigned, the load on $M_2$ is at most $K_2$ and at least $H_{1,2} + H_{1,3} + K_1$ (else, LPT would assign the 1-job to $M_2$). Thus, $K_2 \geq H_{1,2} + H_{1,3} + K_1$. By the flower structure, some job is migrating from $M_1$ to $M_2$. Such a migration is beneficial only if $L'_2 < L_1$. Distinguish between two cases:

1. The 1-job migrates to $M_2$. In this case, $L'_2 = K_2 + H_{1,2} + 1$. Therefore, $K_2 + H_{1,2} + 1 < H_{1,2} + H_{1,3} + K_1 + 1$, or $K_2 < H_{1,3} + K_1$. However, by the above, $K_2 \geq H_{1,2} + H_{1,3} + K_1 \geq H_{1,3} + K_1$. A contradiction.

2. The 1-job does not migrate to $M_2$. In this case, $L'_2 = K_2 + H_{1,2}$. Therefore, $K_2 + H_{1,2} < H_{1,2} + H_{1,3} + K_1 + 1$, or $K_2 < H_{1,3} + K_1 + 1$. However, by the above, $K_2 \geq H_{1,2} + H_{1,3} + K_1 \geq 1 + H_{1,3} + K_1$. A contradiction. The last inequality follows from the fact that $H_{1,2}$ is not empty and consists of at least one job at least as big as the smallest job on $M_1$.

□

**Proposition A.2** *Each of* $K_2, K_3$ *is at least* 1.

*Proof*: We first show $K_2 \geq 1$. Assume $K_2 < 1$, it means that when the 1-job is assigned to $M_1$, the load on $M_2$ is composed of jobs that are a subset of $H_{2,1}$ only. Therefore, by the LPT rule, $H_{2,1} \geq K_1 + H_{1,2} + H_{1,3}$. However, by Proposition 2.8, $L'_2 > L_2$, therefore $H_{2,1} < H_{1,2} + 1$. Thus, $K_1 + H_{1,3} < 1$. However, $H_{1,3} \geq 1$. A contradiction. To show $K_3 \geq 1$, note that if $K_3 > 1$ then by a similar argument to the above $H_{3,1} \geq H_{1,2} + H_{1,3}$. By Proposition 2.8, $L'_3 > L_3$. Therefore $H_{1,3} > H_{3,1}$, implying $K_1 + H_{1,2} < 0$. A contradiction.
□

**Theorem 3.4** *Any LPT configuration on three machines is a* $(\frac{1}{2} + \frac{\sqrt{6}}{4} \approx 1.1123)$*-SE.*
*Proof*: Let $M_1$ be the most loaded machine in the schedule. Recall that the lightest (also last) job assigned to $M_1$ is a 1-job having load 1. Let $\ell = L_1 - 1$. For a give LPT schedule $s$ and a deviation $s' = (s'_\Gamma, s_{-\Gamma})$, let $r = IR_{min}(s, s')$.

By Claim 2.6, $\Gamma$ obeys the flower structure. Therefore: $(i)$ $r \leq L_2/L'_1$; $(ii)$ $r \leq L_3/L'_1$; $(iii)$ $r \leq L_1/L'_2$; and $(iv)$ $r \leq L_1/L'_3$. Let $P = \sum_j p_j$, Clearly, $P = L_1 + L_2 + L_3 = L'_1 + L'_2 + L'_3$. Summing up $(i)$ and $(ii)$, we get

$$L'_1 \leq \frac{L_2 + L_3}{2r} = \frac{P - (\ell + 1)}{2r}. \tag{1}$$

By LPT, $L_2, L_3 \geq \ell$, thus $P \geq 3\ell + 1$. Summing up (*iii*) and (*iv*), and using Equation (1) we get

$$r \leq \frac{2L_1}{L'_2 + L'_3} = \frac{2(\ell+1)}{P - L'_1} \leq \frac{2(\ell+1)}{P(1-\frac{1}{2r}) + \frac{\ell+1}{2r}} \leq \frac{2(\ell+1)}{(3\ell+1)(1-\frac{1}{2r}) + \frac{\ell+1}{2r}}.$$

Implying,

$$r(3\ell+1) - \frac{3\ell+1}{2} + \frac{\ell+1}{2} \leq 2\ell + 2$$

and,

$$r \leq \frac{3\ell+2}{3\ell+1}. \tag{2}$$

**Case 1:** $\ell \geq 3$. In this case, Equation (2) implies $r \leq 1.1$.
**Case 2:** $\ell < 3$. This case requires a closer analysis. Let $I$ be an instance for which LPT creates a schedule with a deviation $s' = (s'_\Gamma, s_{-\Gamma})$ achieving the maximal $IR_{min}$ and $\ell < 3$. For $i = 2, 3$, denote by $H_i$ the total load of jobs migrating from $M_i$ to $M_1$, and by $K_i$ the total load of jobs that remain on $M_i$. By the flower structure, $L'_1 \geq H_2 + H_3$, therefore $H_2 < K_3$ and $H_3 < K_2$, else it would not be beneficial for the jobs composing $H_2, H_3$ to join the coalition. By Propositions A.1 and A.2, each of $H_2, H_3, K_2, K_3$ is at least 1.

**Claim A.3** *The load $\ell$ on $M_1$ is incurred by exactly two jobs.*

*Proof*: Clearly, since we consider the case $\ell < 3$ and the lightest job on $M_1$ has load 1, the load $\ell$ is incurred by at most two jobs. Assume for contradiction that $\ell$ consists of a single job. Then, there are exactly two jobs on $M_1$, of loads $\ell$ and 1. By the flower structure, the $\ell$-job must join the coalition. W.l.o.g assume it migrates to $M_2$. This migration is profitable only if $K_2 < 1$, contradicting Proposition A.2. □

Therefore, we can assume w.l.o.g that in the instance achieving the maximal $IR_{min}$, $M_1$ is assigned three jobs of loads $1+\alpha, 1+\gamma, 1$, for $\alpha, \gamma \geq 0$.

Having $\ell = 2 + \alpha + \gamma$, the bound in Equation (2) implies

$$r \leq \frac{3\ell+2}{3\ell+1} = \frac{8 + 3(\alpha+\gamma)}{7 + 3(\alpha+\gamma)}. \tag{3}$$

Consider first the case in which one of the two big jobs on $M_1$ does not migrate away from $M_1$. We show that no coalition deviation is beneficial in this case. W.l.o.g, assume that the job of length $1+\alpha$ remains on $M_1$ and the job of length $1+\gamma$ migrates to $M_2$. The migration of $1+\gamma$ is profitable only if $K_2 < 2+\alpha$. On the other hand, the migration of the jobs migrating from $M_2$ to $M_1$ is profitable only if $K_2 > H_3 + 1 + \alpha \geq 2 + \alpha$. A contradiction.

Consider next the case in which the 1-job does not migrate away from $M_1$. W.l.o.g, assume that the job of length $1+\gamma$ migrates to $M_2$ and the job of length $1+\alpha$ migrates to $M_3$. In order to bound $r$ according to Equation (3), we find a lower bound for $(\alpha+\gamma)$. By Equation (1),

$$2r \leq \frac{L_2 + L_3}{L'_1} = \frac{K_2 + H_2 + K_3 + H_3}{1 + H_2 + H_3} \leq \frac{K_2 + K_3 + 2}{3} \leq \frac{6 + \alpha + \gamma}{3}. \tag{4}$$

The third inequality is due to the fact that the ratio is decreasing with $H_2 + H_3$, which is known to be at least 2 by Proposition A.1. The last inequality is since the migrations are beneficial for the jobs leaving $M_1$, that is, $K_2 < 2 + \alpha$, and $K_3 < 2 + \gamma$.

Equation (4) implies $6r < 6 + \alpha + \gamma$ or $\alpha + \gamma > 6r - 6$. Next, we apply this bound on $\alpha + \gamma$ into Equation (3) and obtain

$$r < \frac{18r - 10}{18r - 11}.$$

This implies $r < \frac{10}{9} < \frac{1}{2} + \frac{\sqrt{6}}{4}$.

The case we did not analyze yet is the one in which all three jobs assigned to $M_1$ migrate away from $M_1$ in the deviation. Assume w.l.o.g that the jobs of size $1+\gamma$ and 1 are migrating to $M_2$ and the job of size $1 + \alpha$ is migrating to $M_3$. Clearly, the jobs of size $1 + \alpha, 1 + \gamma$ do not migrate to the same machine because they are currently assigned with additional load of 1 while both $K_2$ and $K_3$ are at least 1, by Proposition A.2. Figure 8 shows the schedule before the migration (Figure 8(a)) and after the migration (Figure 8(b)).

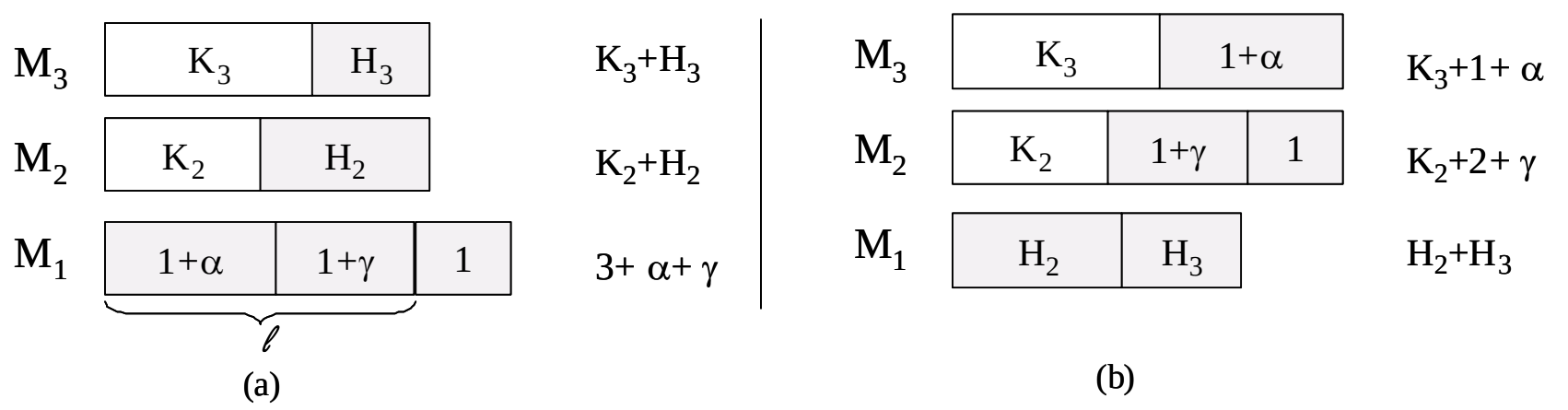

Figure 8: An LPT coalition achieving maximal $IR_{min}$.

Next, we find a lower bound for $\alpha + \gamma$. Considering the migration from $M_1$ to $M_2$, we know that $r \leq (3 + \alpha + \gamma)/(2 + \gamma + K_2)$. Therefore $\alpha + \gamma \geq 2r + \gamma r + K_2 r - 3 \geq 2r + rK_2 - 3$ (because $\gamma \geq 0$). Considering the migration from $M_2$ to $M_1$, we know that $r \leq (K_2 + H_2)/(H_2 + H_3)$. Therefore, $K_2 \geq H_2(r - 1) + H_3 r$. LPT assigns the 1-job on $M_1$ with load $2 + \alpha + \gamma$, while the load on $M_2$ at that time was at most $K_2 + H_2$. Therefore $2 + \alpha + \gamma \leq K_2 + H_2$, implying $H_2 \geq 2 + \alpha + \gamma - K_2$. Also, by Proposition A.1, $H_3 \geq 1$. We can now use these bounds on $H_2, H_3$ to get improved bound on $K_2$. Specifically, $K_2 \geq (2 + \alpha + \gamma - K_2)(r - 1) + r$. This implies $K_2 r \geq 3r + r(\alpha + \gamma) - (2 + (\alpha + \gamma))$. Back to the bound of $\alpha + \gamma$, we now have $\alpha + \gamma \geq 2r + 3r + r(\alpha + \gamma) - (2 + (\alpha + \gamma)) - 3$. Thus, $\alpha + \gamma \geq (5r - 5)/(2 - r)$. Note that $(2 - r)$ is positive since by Theorem 3.1, $r < 5/4$. Finally, we apply this bound on $\alpha + \gamma$ into Equation (3) and obtain

$$r \leq \frac{8 + 3(5r - 5)/(2 - r)}{7 + 3(5r - 5)/(2 - r)} = \frac{1 + 7r}{-1 + 8r}.$$

This implies $r \leq \frac{1}{2} + \frac{\sqrt{6}}{4}$. □

The above bound is tight. Specifically,

**Theorem 3.5** *For any $m \geq 3$, there exists an LPT schedule $s$ such that $IR_{min}(s) = \frac{1}{2} + \frac{\sqrt{6}}{4}$.*

*Proof*: Let $r = \frac{1}{2} + \frac{\sqrt{6}}{4}$, and consider Figure 8, where we substitute $\alpha = \frac{-10+5\sqrt{6}}{6-\sqrt{6}}$, $\gamma = 0$, $K_2 = \frac{r(3+\alpha)-2-\alpha}{r}$, $H_2 = 2 + \alpha - K_2$, $K_3 = 1 + \alpha$, and $H_3 = 1$ (the instance with the rounded values appears in Figure 9). It is easy to verify that all three jobs leaving $M_1$ have improvement ratio of exactly $r = \frac{1}{2} + \frac{\sqrt{6}}{4}$, and the same holds for the two jobs migrating to $M_1$. Thus, in this instance $IR_{min} = \frac{1}{2} + \frac{\sqrt{6}}{4}$. Moreover, this lower bound can be easily extended to any $m > 3$ by adding dummy jobs and machines. Thus,

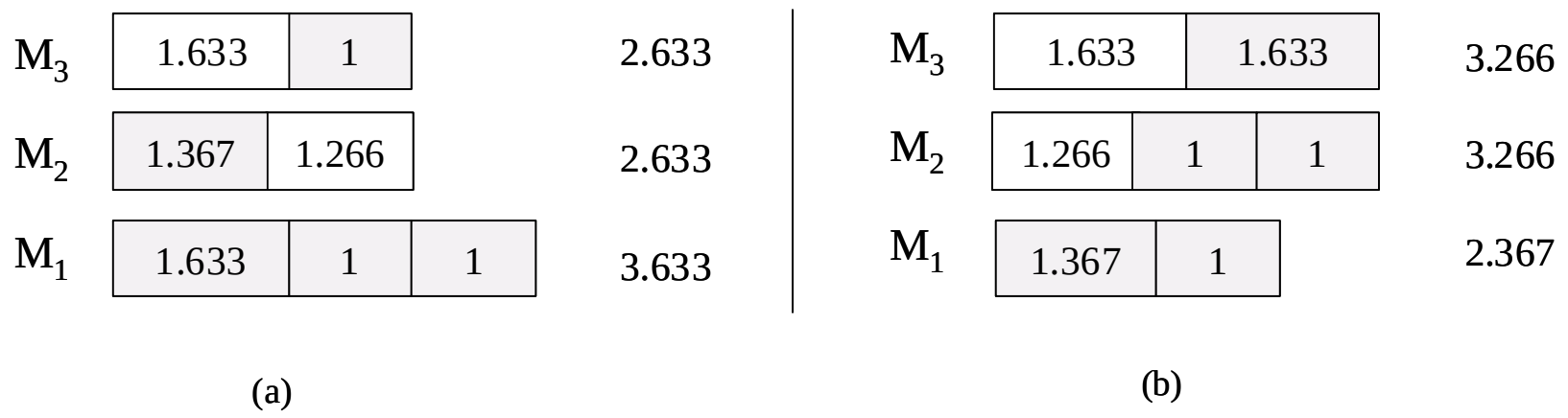

Figure 9: An LPT-originated deviation on three machines in which all migrating jobs improve by $\frac{1}{2} + \frac{\sqrt{6}}{4}$.

□

**Theorem 4.2** *Let s be an LPT schedule on three machines. It holds that* $IR_{max}(s) \leq \frac{5}{3}$.
*Proof*: Let $M_1$ be the most loaded machine. Recall that the lightest (also last) job assigned to $M_1$ is a 1-job having load 1. For $i = 2, 3$, $K_i$ is the set (and also the total load) of jobs that remain on $M_i$, and $H_{i,j}$ is the set (and also total load) of jobs migrating from $M_i$ to $M_j$. For $i = 1$, we let $K_1, H_{1,2}, H_{1,3}$ be as above, but excluding the 1-job.

The 1-job is assigned to $M_1$ by LPT, meaning that the load on $M_2$ and $M_3$ is at least $K_1 + H_{1,2} + H_{1,3}$ at that time. Since the load on $M_2, M_3$ could only increase after the time the 1-job is assigned, we get that

$$K_1 + H_{1,2} + H_{1,3} \leq K_2 + H_{2,1} \quad \text{and} \quad K_1 + H_{1,2} + H_{1,3} \leq K_3 + H_{3,1}. \tag{5}$$

Therefore (sum up the two):

$$2(K_1 + H_{1,2} + H_{1,3}) \leq K_2 + K_3 + H_{2,1} + H_{3,1}. \tag{6}$$

Distinguish between two cases:

**(i) The 1-job remains on $M_1$.** In This case, $L_1 = K_1 + H_{1,2} + H_{1,3} + 1$; $L_2 = K_2 + H_{2,1}$; $L_3 = K_3 + H_{3,1}$, while after the coalition is active $L'_1 = K_1 + H_{2,1} + H_{3,1} + 1$; $L'_2 = K_2 + H_{1,2}$; $L'_3 = K_3 + H_{1,3}$.

Since the jobs in $H_{1,2}$ and $H_{1,3}$ are part of the coalition, $L'_2 + L'_3 < 2L_1$. Deducing $H_{1,2}$ and $H_{1,3}$ from both sides we get $K_2 + K_3 < H_{1,2} + H_{1,3} + 2K_1 + 2$. Combining with Equation 6, we get:

$$H_{1,2} + H_{1,3} < H_{2,1} + H_{3,1} + 2. \tag{7}$$

By Proposition A.1, each of $H_{2,1}, H_{3,1}, K_2, K_3$ is at least 1. By Proposition 2.9, the improvement ratio of the 1-job, which equals $L_1/L'_1$, is the largest among the coalition. This ratio can now be bounded as follows:

$$\frac{L_1}{L'_1} = \frac{K_1 + H_{1,2} + H_{1,3} + 1}{K_1 + H_{2,1} + H_{3,1} + 1} < \frac{K_1 + H_{2,1} + H_{3,1} + 3}{K_1 + H_{2,1} + H_{3,1} + 1} \leq \frac{5}{3}.$$

The left inequality follows from Equation 7. The right one follow from Proposition A.1 and from the fact that $K_1$ might be empty.

**(ii) The 1-job leaves $M_1$.** We assume w.l.o.g that the 1-job moves to $M_2$. In This case, $L_1 = K_1 + H_{1,2} + H_{1,3} + 1$; $L_2 = K_2 + H_{2,1}$; $L_3 = K_3 + H_{3,1}$, while after the coalition is active $L'_1 = K_1 + H_{2,1} + H_{3,1}$; $L'_2 = K_2 + H_{1,2} + 1$; $L'_3 = K_3 + H_{1,3}$.

Since the jobs in $H_{1,2}$ and $H_{1,3}$ are part of the coalition, $L'_2 + L'_3 < 2L_1$. Deducing $1, H_{1,2}$ and $H_{1,3}$ from both sides we get $K_2 + K_3 < H_{1,2} + H_{1,3} + 2K_1 + 1$. Combining with Equation 6, we get:

$$H_{1,2} + H_{1,3} < H_{2,1} + H_{3,1} + 1. \tag{8}$$

By Propositions A.1 A.2, each of $H_{2,1}, H_{3,1}, K_2, K_3$ is at least 1. If $K_1$ is not empty then the jobs of $K_1$ have improvement ratio $L_1/L'_1$ which is, by Proposition 2.9, the largest ratio among the coalition. This ratio can now be bounded as follows:

$$\frac{L_1}{L'_1} = \frac{K_1 + H_{1,2} + H_{1,3} + 1}{K_1 + H_{2,1} + H_{3,1}} \leq \frac{K_1 + H_{2,1} + H_{3,1} + 2}{K_1 + H_{2,1} + H_{3,1}} < \frac{5}{3}.$$

The left inequality follows from Equation 8. The right one follows from Proposition A.1, and from the fact that $K_1$ is not empty and includes at least one job of load at least 1.

If $K_1$ is empty, then as we show below, the maximal improvement ratio is less than 3/2. We bound separately the improvement ratio of $H_{1,2}, H_{1,3}$, and $H_{i,1} (i \in \{1, 2\})$. Denote by $r_{i,j}$ the IR of jobs moving from $M_i$ to $M_j$. In addition to Equations 5 and 8, and to Propositions A.1 and A.2, we also use below Proposition 2.8. Specifically, $H_{2,1} < H_{1,2} + 1$ and $H_{3,1} < H_{1,3}$. Finally, bear in mind that $K_1 = \emptyset$.

$$r_{1,2} = \frac{L_1}{L'_2} = \frac{H_{1,2} + H_{1,3} + 1}{K_2 + H_{1,2} + 1} \leq \frac{K_2 + H_{2,1} + 1}{K_2 + H_{1,2} + 1} < \frac{K_2 + H_{2,1} + 1}{K_2 + H_{2,1}} < \frac{3}{2}.$$

$$r_{1,3} = \frac{L_1}{L'_3} = \frac{H_{1,2} + H_{1,3} + 1}{K_3 + H_{1,3}} \leq \frac{K_3 + H_{3,1} + 1}{K_3 + H_{1,3}} < \frac{K_3 + H_{3,1} + 1}{K_3 + H_{3,1}} < \frac{3}{2}.$$

$$r_{i,1} = \frac{L_i}{L'_1} = \frac{K_i + H_{i,1}}{H_{2,1} + H_{3,1}} < \frac{H_{1,2} + H_{1,3}}{H_{2,1} + H_{3,1}} < \frac{H_{2,1} + H_{3,1} + 1}{H_{2,1} + H_{3,1}} < \frac{3}{2}.$$

□

## Appendix B. List Scheduling

List Scheduling (LS) is a greedy scheduling algorithms in which the jobs are assigned to the machines in arbitrary order, but similar to LPT, each job is assigned to the least loaded machine at the time of assignment. LS is known to provide a $(2 - \frac{1}{m})$-approximation to the minimum makespan (Graham, 1966). While LS does not depart qualitatively from LPT with respect to makespan approximation (i.e., both provide a constant approximation to the optimal makespan), they are qualitatively different with respect to game theoretic properties. First, LS does not necessarily produce an NE. Moreover, as we next show, LS performs poorly with respect to the measures introduced in this paper.

The improvement ratio of a job is not bounded even if the coalition consists of a single job. Consider for example an instance with 2 machines and jobs of lengths $1, 1, X$ (in that order) such that $X > 1$. LS will produce a schedule with loads $1, 1 + X$. The job of length 1 scheduled with the long job can migrate and join the other short job. Its improvement ratio is $1 + X/2$ which can be arbitrarily large.

The damage ratio of a deviation from an LS schedule is not bounded either. Consider an instance with three machines and jobs of lengths $\{1 - 2\varepsilon, 1 - \varepsilon, 1, 2, X, 2, 3\}$. It is easy to verify that in the resulting LS-configuration, there exists a coalition in which the job of length $X$ migrates. Since $X$ can be arbitrarily large, the damage ratio of the job in the machine into which $X$ migrates is arbitrarily large. We note that the damage ratio caused by a deviation of a single job is at most 2. To see this, consider an LS configuration and assume that a job $j$ of length $p_j$ migrates from $M_1$ to $M_2$. Denote by $B_j, A_j$ the total load of jobs on $M_1$ that were assigned before and after $j$ respectively. If $A_j = 0$ ($j$ is last) then it is not beneficial for $j$ to migrate ($B_j < L_2$, else $j$ should have been assigned to $M_2$). Else, the first job after $j$ was assigned to $M_1$ because $B_j + p_j$ was less than the load at that time on $M_2$. Therefore $L_2 \geq B_j + p_j$, and in particular $p_j \leq L_2$. The damage ratio is $(L_1 + p_j)/L_1 \leq 2$. The analysis is tight as can be exemplified by the instance $m = 2, I = \{1, 1, X\}$.